\documentclass[journal,twoside,web]{ieeecolor}
\usepackage{tmi}
\usepackage{cite}
\usepackage{amsmath,amssymb,amsfonts}
\usepackage{algorithmic}
\usepackage{graphicx}
\usepackage{booktabs}
\usepackage{caption}
\usepackage{subcaption}
\usepackage{multirow}
\usepackage{textcomp}
\usepackage[switch]{lineno}
\newcommand{\tabincell}[2]{\begin{tabular}{@{}#1@{}}#2\end{tabular}}
\def\BibTeX{{\rm B\kern-.05em{\sc i\kern-.025em b}\kern-.08em
    T\kern-.1667em\lower.7ex\hbox{E}\kern-.125emX}}
    
\begin{document}
\title{Adaptive Hierarchical Dual Consistency for Semi-Supervised Left Atrium Segmentation on Cross-Domain Data}

\author{Jun Chen, Heye Zhang, Member, IEEE, Raad Mohiaddin, Tom Wong, David Firmin, Jennifer Keegan, and Guang Yang, Senior Member, IEEE

\thanks{This study was supported in part by the Key-Area Research and Development Program of Guangdong Province (2019B010110001), the Key Program for International Cooperation Projects of Guangdong Province (2018A050506031), the National Youth Talent Support Program (RC2020-01), the Guangdong Natural Science Funds for Distinguished Young Scholar (2019B151502031), the Natural Science Foundation of Guangdong Province (2020B1515120061), and the National Natural Science Foundation of China (61771464, U1801265, U1908211); in part by the British Heart Foundation (Project Number: TG/18/5/34111, PG/16/78/32402), the European Research Council Innovative Medicines Initiative (DRAGON, H2020-JTI-IMI2 101005122), the AI for Health Imaging Award (CHAIMELEON, H2020-SC1-FA-DTS-2019-1 952172), and the UK Research and Innovation Future Leaders Fellowship (MR/V023799/1).}
\thanks{H. Zhang and G. Yang are corresponding authors (e-mail: zhangheye@mail.sysu.edu.cn; g.yang@imperial.ac.uk).}
\thanks{J. Chen and H. Zhang are with the School of Biomedical Engineering, Sun Yat-sen University, Shenzhen, Guangdong 518107, P.R. China.}
\thanks{R. Mohiaddin, T. Wong, D. Firmin, J. Keegan and G. Yang are with the Cardiovascular Research Centre, Royal Brompton Hospital, SW3 6NP, London, U.K and the National Heart and Lung Institute, Imperial College London, London, SW7 2AZ, U.K.}
}

\maketitle

\begin{abstract}
Semi-supervised learning provides great significance in left atrium (LA) segmentation model learning with insufficient labelled data. Generalising semi-supervised learning to cross-domain data is of high importance to further improve model robustness. However, the widely existing distribution difference and sample mismatch between different data domains hinder the generalisation of semi-supervised learning. In this study, we alleviate these problems by proposing an \emph{Adaptive Hierarchical Dual Consistency} (AHDC) for the semi-supervised LA segmentation on cross-domain data. The AHDC mainly consists of a Bidirectional Adversarial Inference module (BAI) and a Hierarchical Dual Consistency learning module (HDC). The BAI overcomes the difference of distributions and the sample mismatch between two different domains. It mainly learns two mapping networks adversarially to obtain two matched domains through mutual adaptation. The HDC investigates a hierarchical dual learning paradigm for cross-domain semi-supervised segmentation based on the obtained matched domains. It mainly builds two dual-modelling networks for mining the complementary information in both intra-domain and inter-domain. For the intra-domain learning, a consistency constraint is applied to the dual-modelling targets to exploit the complementary modelling information. For the inter-domain learning, a consistency constraint is applied to the LAs modelled by two dual-modelling networks to exploit  the  complementary knowledge among different data domains. We demonstrated  the  performance  of  our  proposed  AHDC  on  four 3D late gadolinium enhancement cardiac MR (LGE-CMR) datasets from different centres and a 3D CT dataset. Compared to other state-of-the-art methods, our proposed AHDC achieved higher segmentation accuracy, which indicated its capability in the cross-domain semi-supervised LA segmentation.
\end{abstract}


\section{Introduction}
\label{sec:introduction}
\IEEEPARstart{S}{emi-supervised} learning provides great significance in left atrium (LA) segmentation model learning with insufficient labelled data. Automated and accurate LA segmentation is a crucial task to aid the diagnosis and treatment for the patients with atrial fibrillation (AF) \cite{razeghi2020fully,xiong2019fully,yang2020simultaneous,chen2021jas}. Deep learning based approaches have great potential for the LA segmentation \cite{zhang2021fully,xiong2020global}. However, it is expensive and laborious to annotate large amounts of data  by experienced experts  for training an accurate LA segmentation model based on deep learning \cite{yu2019uncertainty}. Since semi-supervised learning can alleviate the need for the labelled data by effectively exploiting the unlabelled data to learn deep models \cite{cao2021uncertainty}. Semi-supervised learning is able to overcome the insufficient labelled data for advancing the accurate LA segmentation, benefiting the subsequent diagnosis and treatment for the patients with AF.

Generalising semi-supervised learning to cross-domain data for the LA segmentation is of high importance to improve model robustness. Semi-supervised learning aims to mine effective hidden information from unlabelled data to support model learning \cite{chapelle2009semi}. Because of the noise interference and the limited collection capabilities of data sources, a single data domain can not always provide sufficient high-quality unlabelled data and abundant data characteristics for robust semi-supervised LA segmentation. For example, the single data domain is usually subject to the limited LA varieties of contrast, shape and texture for robust model learning. Compared to the single data domain, cross-domain data not only can provide more available high-quality data, but also can provide complementary domain information and more comprehensive data characteristics to describe the LA of interest \cite{yang2019comprehensive}. Therefore, it is important to effectively ensemble cross-domain data for robust semi-supervised LA segmentation. 

\begin{figure}[!hbtp]
\begin{center}
\scalebox{.99}{
   \includegraphics[width=1\linewidth]{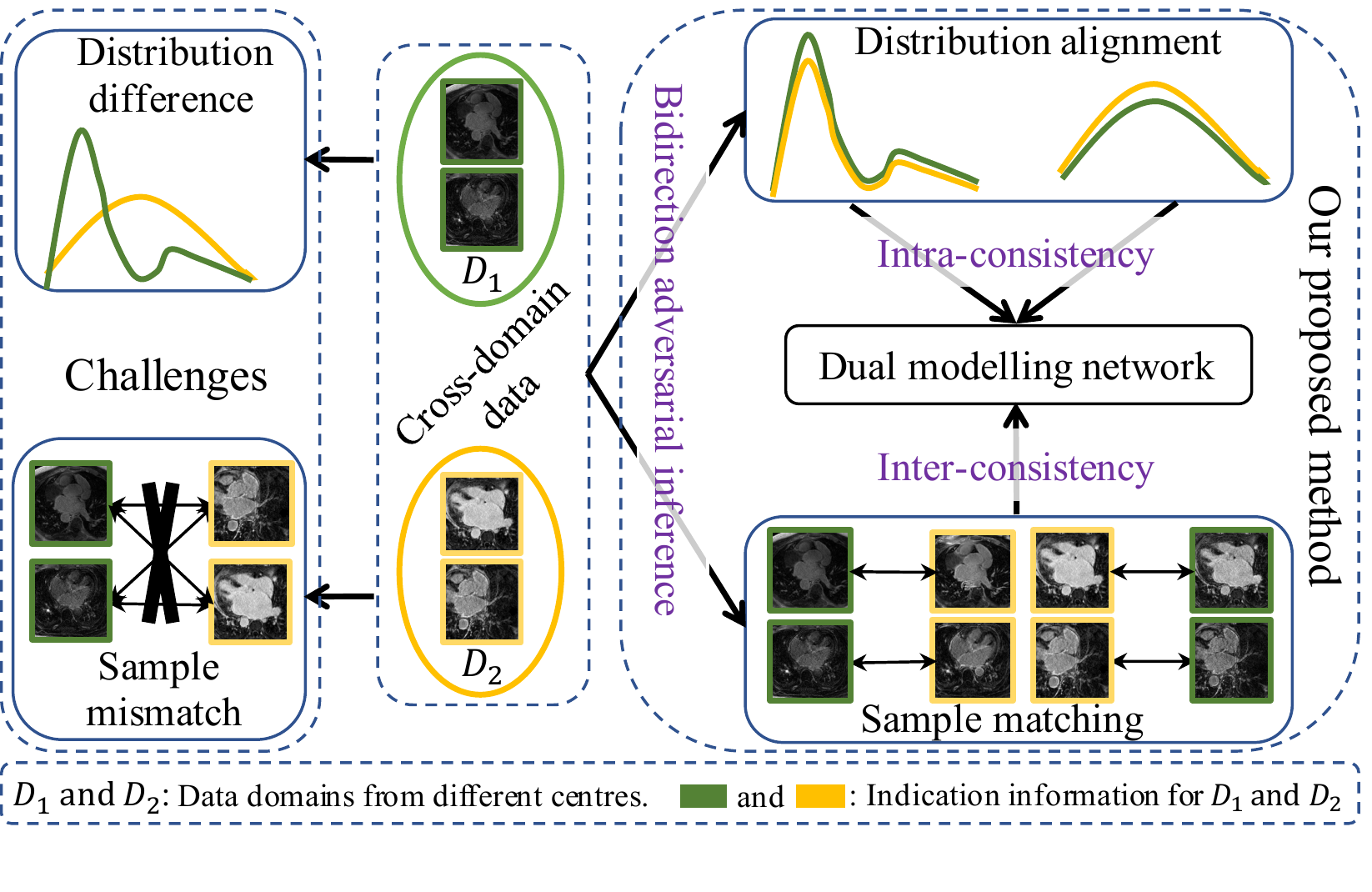}
   }
\end{center}
\caption{Our proposed adaptive hierarchical dual consistency overcomes the difference of data distribution and sample mismatch in different domains for the cross-domain semi-supervised segmentation.}
\label{fig:ahdc_v} 
\end{figure}

However, generalising semi-supervised to cross-domain data is difficult due to the difference of distributions and the sample mismatch as shown in Fig. \ref{fig:ahdc_v}: (1) The difference of cross-domain data distributions. Semi-supervised learning with the generative model, low-density separation and graph-based method can work but relies on the consistent data distribution under certain model assumptions including smoothness assumption, cluster assumption or manifold assumption \cite{chapelle2009semi}. Performance degradation of the semi-supervised model may occur whenever the assumptions adopted for a particular task do not match the characteristics of the data distribution \cite{chapelle2009semi}. In the real world, cross-domain data collected from different sources exhibit heterogeneous properties \cite{campello2021multi}, which can lead to the difference in distributions. For example, in medical image analysis, because of the different subject groups, scanners, or scanning protocols, the distributions of cross-domain data are different \cite{cheplygina2018transfer}. Therefore, generalising semi-supervised learning to cross-domain data directly is not trivial. (2) Sample mismatch of cross-domain data. Semi-supervised learning with the disagreement-based method requires matched samples from different domains, where the information of different domains is regarded as the different characteristics of matched samples \cite{dong2018tri}. Since the collection of cross-domain data is independent, the samples in different domains are not matched. This restricts the cross-domain generalisation of semi-supervised learning.

In order to overcome the issues mentioned above, we propose an \emph{\textbf{A}daptive \textbf{H}ierarchical \textbf{D}ual \textbf{C}onsistency} framework called \textbf{AHDC} for semi-supervised  LA segmentation on cross-domain data as shown in Fig. \ref{fig:ahdc_v}. The AHDC consists of two modules: (1) A Bidirectional Adversarial Inference module (BAI), which performs the mutual domain adaptation to align distributions and match samples for two different data domains. The adapted domains and two corresponding source domains are merged to obtain two matched domains. The obtained matched domains not only expand the number of data in a specific source domain, but also learns complementary representation for the samples in the specific source domain. (2) A Hierarchical Dual Consistency learning module (HDC), which performs a hierarchical semi-supervised segmentation with dual consistency on the obtained matched domains. The HDC builds two dual-modelling networks applied to the matched domains for mining the complementary information in both intra-domain and inter-domain. Within a specific domain, the segmentation task is represented as global modelling and local modelling. Then we perform a consistency between the complementary modelling LAs for intra-domain semi-supervised learning. For the inter-domain,  we build a consistency between the outputs of dual-modelling networks estimated from different domains to exploit the complementary domain information.

Our main contributions are summarised as follows:
\begin{itemize}
\item We propose a semi-supervised LA segmentation framework for generalising across domains. It provides a solution for generalising semi-supervised LA segmentation to cross-domain data with effectiveness on both different distributions and mismatched samples. 

\item We propose a paradigm of hierarchical dual consistency learning to mine the effective information in both inter-domain and intra-domain. It explicitly enforcing consistency under complementary information.

\item We have conducted comprehensive experiments on four 3D MR datasets from different centres and one 3D CT dataset. The experiment results demonstrated the feasibility and the superiority of our proposed cross-domain semi-supervised segmentation framework. 
\end{itemize}

\section{Related Work}
\subsection{Domain Adaptation}
Domain adaptation, which aims to overcome the distribution difference of different domains, has drawn great attention in computer vision \cite{patel2015visual}. Because generative adversarial network (GAN) has great superiority in capturing data distribution, it has been widely used in domain adaptation for aligning distributions of different domains \cite{li2020towards,zhang2020collaborative, chen2019discriminative, chen2020unsupervised, chen2019joint}. There are different GAN based structures for achieving  domain adaptation. For the domain adaptation with a single direction, GAN usually leverages a generator and a discriminator to improve the distribution of the source domain to approximate it to the distribution of the target domain by adversarial learning. To focus on the high-resolution image with emphasis on pixel-level reconstruction, Pix2pixHD extends conditional GANs to leverage a decomposed generator and three multi-scale discriminators to achieve domain adaptation \cite{wang2018high}. For the domain adaptation with bi-direction, CycleGAN \cite{zhu2017unpaired}, DualGAN \cite{yi2017dualgan} and DiscoGAN \cite{kim2017learning} concatenate two generators with two discriminators to ensure two cyclic consistency for the bidirectional domain adaptation of two different domains. ALI \cite{dumoulin2017adversarially} and BiGAN \cite{donahue2017adversarial} employ two generators and a discriminator to match joint distribution for different domains. However, ALI and BiGAN do not focus on pixel-level reconstruction, thus cannot effectively capture the position, colour, and style of targets. ALICE extends the ALI to exploit cycle-consistency to focus on pixel-level reconstruction for the target domain \cite{li2017alice}. It also proposes to enforce cycle-consistency using fully adversarial learning with an extra discriminator. Our used domain adaptation method is based on the ALICE framework. We extended it to focus on bidirectional pixel-level reconstruction for two domains simultaneously. In order to reduce computing resources and difficulty of training while using fully adversarial learning, we adopt the explicit cycle-consistency, thus exploiting two generators and a discriminator for bidirectional domain adaptation with pixel-level reconstruction.

\subsection{Semi-supervised Learning} 
Semi-supervised learning alleviates the problem of the lack of labelled data. Here we only discuss related consistency-based and disagreement-based semi-supervised learning. More information about semi-supervised learning can be found in \cite{chapelle2009semi}. The consistency-based methods constrain the prediction consistency under different perturbations and ensembles. For example, the $\Pi$ model enforces the prediction consistency under the input perturbations with different Gaussian noise and the model perturbation with dropout operation \cite{samuli2017temporal}. Unsupervised data augmentation (UDA) replaces the traditional noise perturbations with high-quality data augmentations (e.g., RandAugment, Back-translation and TF-IDF) to improve consistency learning \cite{xie2020unsupervised}. FixMatch uses a separate weak augmentation and a strong augmentation on input data for consistency regularisation \cite{sohn2020fixmatch}. In contrast to these methods, Temporal Ensembling (TE) penalises the inconsistency between the current prediction and the integration of previous predictions based on an exponential moving average (EMA) \cite{samuli2017temporal}. Compared to the TE, the Mean Teacher proposes to average the weights of a base model \cite{tarvainen2017mean}. However, they need multiple reasoning processes  to provide predictions for consistency learning, thus being subject to the computational cost. 

The disagreement-based semi-supervised learning exploits the disagreement of predictions from multiple task learners during the learning process \cite{dong2018tri} including co-training and co-regularisation. Co-training leverages two sufficient and redundant views of data to train two task models for annotating the unlabelled data. Then the unlabelled data with high prediction confidence is added to the training set for further improving the model \cite{qiao2018deep,xia2020uncertainty}. Co-regularisation tries to directly minimise the prediction disagreement of unlabelled samples on different views \cite{zhao2017multi}. 
\begin{figure*}[!ht]
	\begin{center}
	\scalebox{.92}{
	 \includegraphics[width=1\textwidth]{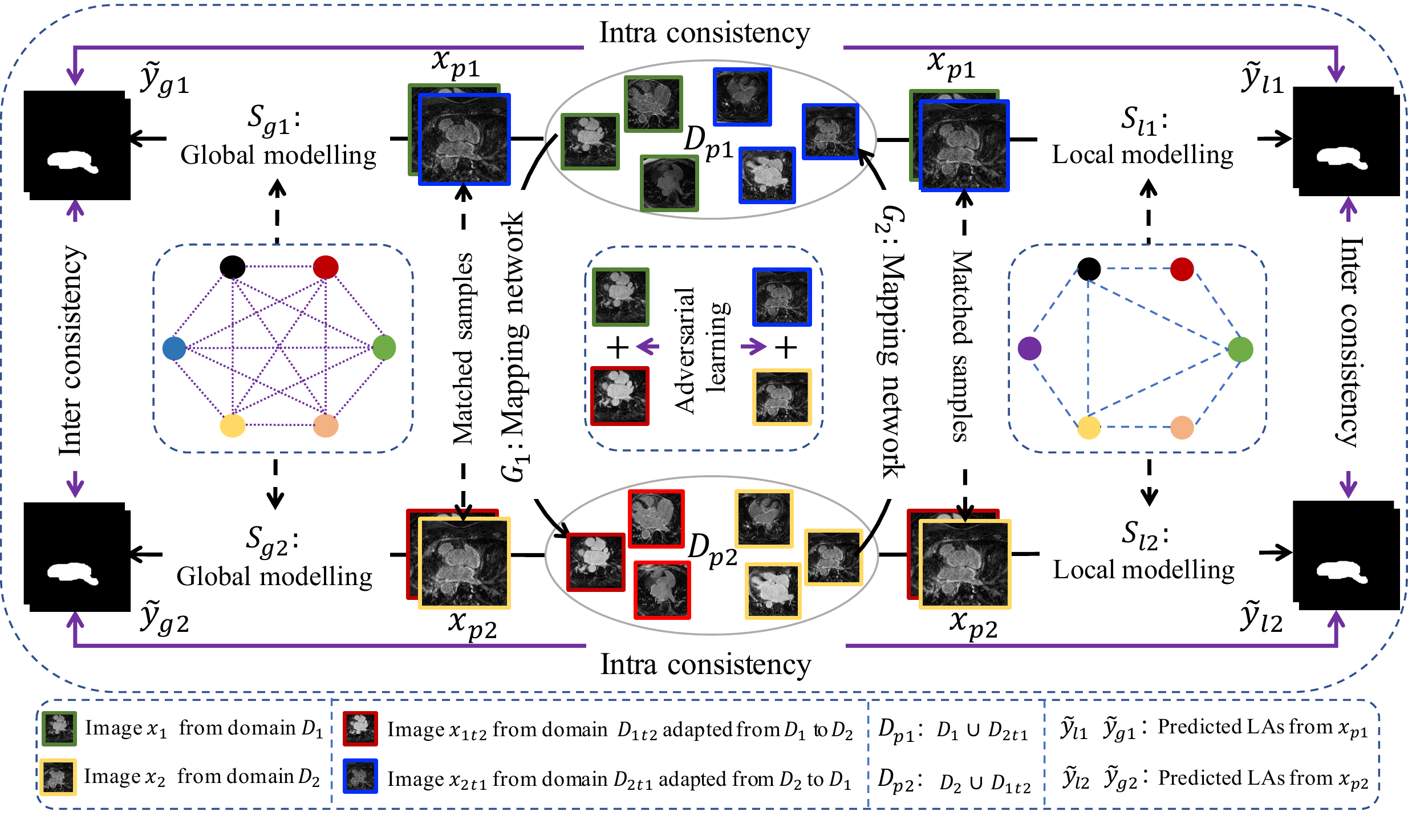}
	 }
	\end{center}
	\caption{Overview of our proposed AHDC framework for cross-domain semi-supervised segmentation. The framework consists of a bidirectional adversarial inference (BAI) module and a hierarchical dual consistency learning (HDC) module. The BAI module employs two mapping networks to perform a mutual adaptation of two different domains of $D_{1}$ and $D_{2}$ to obtain matched domains of $D_{p1}$ and $D_{p2}$. The HDC module applies two dual-modelling networks to the matched domains for performing semi-supervised segmentation tasks. Each dual-modelling network contains a global-modelling branch ($S_{g1}$/$S_{g2}$) used to capture the global correlation of feature maps to estimate LA ($\widetilde{y}_{g1}/\widetilde{y}_{g2}$), and a local-modelling branch ($S_{l1}$/$S_{l2}$) used to capture the local correlation of feature maps to estimate LA ($\widetilde{y}_{l1}/\widetilde{y}_{l2}$). In intra-domain, a consistency is performed between $\widetilde{y}_{l1}/\widetilde{y}_{l2}$ and $\widetilde{y}_{g1}/\widetilde{y}_{g2}$ estimated by complementary modellings, respectively. In inter-domain, a consistency is performed between $\widetilde{y}_{l1}/\widetilde{y}_{g1}$ and $\widetilde{y}_{l2}/\widetilde{y}_{g2}$ estimated by complementary domain networks, respectively.}\label{fig:AHDC}
\end{figure*}

\begin{table}[!hbtp]
\captionsetup{justification=centering}
 \caption{Summary of notations}
 \centering
 \scalebox{.80}{
    \begin{tabular}{|c|c|c|c|} 
    \hline
    Notion & Definition & Notion & Definition\cr \hline
     $D_{1}$   &Domain from source1  &$D_{2}$   &Domain from source2\cr \hline
     $D_{1t2}$ &\tabincell{c}{Domain adapted\\from $D_{1}$ to $D_{2}$} &$D_{2t1}$ & \tabincell{c}{Domain adapted\\from $D_{2}$ to $D_{1}$}\cr \hline
     $D_{p1}$ &$D_{1}\cup D_{2t1}$ &$D_{p2}$  & $D_{2}\cup D_{1t2}$ \cr \hline
     $D^{l},D^{u}$ &\tabincell{c}{Labelled domain, \\Unlabelled domain}  &$G_{1}, G_{2}$   &\tabincell{c}{Mutual mapping nets \\ of $D_{1}$ and $D_{2}$} \cr \hline 
     \tabincell{c}{$S_{1}=\{S_{l1}, S_{g1}\},$ \\ $S_{2}=\{S_{l2},S_{g2}\}$} & \tabincell{c}{Dual-modelling nets \\ $\{$local net, global net$\}$} &$T$ &Discriminator \cr \hline 
     \tabincell{c}{$x_{1},x_{2},x_{1t2}$,\\$x_{2t1},x_{p1},x_{p2}$} &\tabincell{c}{Images from $D_{1},D_{2}$,\\$D_{1t2},D_{2t1},D_{p1},D_{p2}$}  &\tabincell{c}{$\widetilde{y}_{l1},\widetilde{y}_{g1}$, \\ $\widetilde{y}_{l2},\widetilde{y}_{g2}$} & \tabincell{c}{Estimated LAs from \\  $S_{l1},S_{g1},S_{l2},S_{g2}$}\cr \hline
     $\hat{x}_{1},\hat{x}_{2}$ & \tabincell{c}{Reconstructions\\of $x_{1},x_{2}$} & $x^{l},x^{u}$ &\tabincell{c}{Labelled data,\\unlabelled data} \cr \hline
     $j(x_{p1},x_{p2})$  & \tabincell{c}{Joint distribution \\of $D_{p1}$,$D_{p2}$}  &$y$ &Ground truth \cr \hline
      \tabincell{c}{$p(x_{1}),q(x_{2})$, \\$p(x_{p1}),q(x_{p2})$} &\tabincell{c}{Marginal distributions \\of $D_{1}$,$D_{2}$,$D_{p1}$,$D_{p2}$} &\tabincell{c}{$p_{\varphi_{1}}(x_{2}|x_{1})$ \\ $q_{\varphi_{2}}(x_{1}|x_{2})$}  &\tabincell{c}{Parameterised \\ conditional distributions} \cr \hline
      $\varphi_{1},\varphi_{2}$ &Params of $G_{1},G_{2}$ &$\psi_{1}$ &Param of $T$ \cr \hline
      \tabincell{c}{$\theta_{1}=$ \\ $\{\theta^{f}_{1},\theta^{l}_{1},\theta^{g}_{1}\}$} &\tabincell{c}{Param of $S_{1}$ in the \\ modules of feature, \\ local-modelling, \\ global-modelling} &\tabincell{c}{$\theta_{2}=$ \\ $\{\theta^{f}_{2},\theta^{l}_{2},\theta^{g}_{2}\}$} &\tabincell{c}{Param of $S_{2}$ in the \\ modules of feature, \\ local-modelling, \\ global-modelling} \cr \hline
      $L (\cdot)$ & Loss function &$\lambda$ & Weight Param \cr \hline
    \end{tabular}
 }
 \label{table:correlation results}
\end{table}

\section{Method}
\subsection{Overview}
The overview of our proposed AHDC framework is illustrated in Fig. \ref{fig:AHDC}. The notations are summarised in TABLE \uppercase\expandafter{\romannumeral1}. The AHDC framework consists of two modules: a BAI module and a HDC module. Given two different data domains denoted by $D_{1}$ and $D_{2}$. $D_{1}$ contains both labelled data $D^{l}_{1}$ and unlabelled data $D^{u}_{1}$, where $D^{l}_{1}=\{((x^{l}_{1})^{i},y^{i})\}^{n_{1}}_{i=1}$ with $n_{1}$ labelled samples and  $D^{u}_{1}=\{(x^{u}_{1})^{i}\}^{n_{1}+n_{2}}_{i=n_{1}+1}$ with $n_{2}$ unlabelled samples, respectively. The $D_{2}$ only contains unlabelled data denoted as $D^{u}_{2}=\{(x^{u}_{2})^{i}\}^{m_{1}}_{i=1}$ with $m_{1}$ unlabelled samples. The BAI module employs two mapping networks of $G_{1}$ and $G_{2}$ to generate complementary domains by adapting $D_{1}$ and $D_{2}$ to each other, where the domain adapted from $D_{1}$ to $D_{2}$ is denoted as $D_{1t2}$ while the domain adapted from  $D_{2}$ to $D_{1}$ is denoted as $D_{2t1}$. Then the targeted domains ($D_{1}$ and $D_{2}$) and the corresponding adapted domains ($D_{2t1}$ and $D_{1t2}$) merge to form two matched domains of $D_{p1}$ and $D_{p2}$. Finally, two dual-modelling networks of  $S_{1}=\{S_{l1},S_{g1}\}$ and $S_{2}=\{S_{l2},S_{g2}\}$ are fed with matched samples sampled from $D_{p1}$ and $D_{p2}$ to predict LAs, where the LAs predicted by the local modelling $S_{l1}$ and the global modelling $S_{g1}$ are denoted as $\widetilde{y}_{l1}$ and $\widetilde{y}_{g1}$ while the LAs predicted by the local modelling $S_{l2}$ and the global modelling $S_{g2}$ are denoted as $\widetilde{y}_{l2}$ and $\widetilde{y}_{g2}$, respectively.

\subsection{Bidirectional Adversarial Inference for Distribution Alignment and Sample Matching.}
Consider a $D_{1}$ to $D_{2}$ domain mapping network $G_{1}:x_{1} \rightarrow x_{2}$. Meanwhile, consider a $D_{2}$ to $D_{1}$ domain mapping network $G_{2}:x_{2} \rightarrow x_{1}$. We denote two domain marginal distributions of $D_{1}$ and $D_{2}$ as $p(x_{1})$ and $q(x_{2})$. One domain can be inferred based on the other using parameterised conditional distributions, $p_{\varphi_{1}}(x_{2}|x_{1})$ and $q_{\varphi_{2}}(x_{1}|x_{2})$, where $\varphi_{1}$ and $\varphi_{2}$ denote the parameters of two distributions. Then, we have the joint distributions of $p_{\varphi_{1}}(x_{1},x_{2})=p_{\varphi_{1}}(x_{2}|x_{1})p(x_{1})$ and $q_{\varphi_{2}}(x_{1},x_{2})=q_{\varphi_{2}}(x_{1}|x_{2})q(x_{2})$. We aims to match $p_{\varphi_{1}}(x_{2})=\int p_{\varphi_{1}}(x_{2},x_{1})dx_{1}$ to $q(x_{2})$ and match $q_{\varphi_{2}}(x_{1})=\int q_{\varphi_{2}}(x_{1},x_{2})dx_{2}$ to $p(x_{1})$ by matching $p_{\varphi_{1}}(x_{1},x_{2})$ and $q_{\varphi_{2}}(x_{1},x_{2})$. Then we use a discriminator network $T_{\psi_{1}}(x_{1},x_{2})$ parameterised using $\psi_{1}$ to penalise mismatches in the joint distributions of $p_{\varphi_{1}}(x_{1},x_{2})$ and $q_{\varphi_{2}}(x_{1},x_{2})$. Specifically, we consider following objectives:
\begin{equation}
\begin{split}
&\mathop{\min_{\varphi_{1},\varphi_{2}}\max_{\psi_{1}}} O^{d}(\varphi_{1},\varphi_{2},\psi_{1})\\
&=E_{(x_{1},x_{2})\sim p_{\varphi_{1}}(x_{1},x_{2})}[log\ \sigma(T_{\psi_{1}}(x_{1},x_{2}))]\\
&+E_{(x_{1},x_{2})\sim q_{\varphi_{2}}(x_{1},x_{2})}[1-log\ \sigma( T_{\psi_{1}}(x_{1},x_{2}))]\\
\end{split}
\end{equation}\label{equationali}
where the $\sigma(\cdot)$ denotes the sigmoid function. 

Intuitively, if equation (1) is achieved, $p_{\varphi_{1}}(x_{1},x_{2})$ and $q_{\varphi_{2}}(x_{1},x_{2})$ match each other, which not only implies that $p_{\varphi_{1}}(x_{2})$ and $q(x_{2})$ match each other, but also implies that $q_{\varphi_{2}}(x_{1})$ and $p(x_{1})$ match each other. However, the relationship between random variables $x_{1}$ and $x_{2}$ is not specified or constrained by equation (1). In order to obtain paired samples, according to \cite{li2017alice}, we extend the conditional entropies from single constraint to bi-direction constraints $(H(x_{1}|x_{2})$ and $H(x_{2}|x_{1}))$, which imposes constraints on the conditionals $p_{\varphi_{1}}(x_{2}|x_{1})$ and $q_{\varphi_{2}}(x_{1}|x_{2})$, simultaneously. Because there is no explicit distributions to compute the conditional entropies.  According to \cite{li2017alice}, we bound the conditional entropies using the cycle-consistency ($L^{x_{1}\rightarrow \hat{x}_{1}}(\varphi_{1},\varphi_{2})$ and $L^{x_{2}\rightarrow \hat{x}_{2}}(\varphi_{1},\varphi_{2})$):
\begin{equation}
\begin{split}
&H(x_{1}|x_{2})\\
=&-E_{x_{1}\sim p(x_{1}),x_{2}\sim p_{\varphi_{1}}(x_{2}|x_{1})}[log p_{\varphi_{1}}(x_{1}|x_{2})]\\
=&-E_{x_{1}\sim p(x_{1}),x_{2}\sim p_{\varphi_{1}}(x_{2}|x_{1})}[log q_{\varphi_{2}}(x_{1}|x_{2})]\\
&-E_{x_{1}\sim p(x_{1}),x_{2}\sim p_{\varphi_{1}}(x_{2}|x_{1})}[log p_{\varphi_{1}}(x_{1}|x_{2})-log q_{\varphi_{2}}(x_{1}|x_{2})]\\
=&-E_{x_{1}\sim p(x_{1}),x_{2}\sim p_{\varphi_{1}}(x_{2}|x_{1})}[log q_{\varphi_{2}}(x_{1}|x_{2})]\\
&-E_{q_{\varphi_{2}(x_{2})}}[KL(p_{\varphi_{1}}(x_{1}|x_{2})||q_{\varphi_{2}}(x_{1}|x_{2}))]\\
\leq &-E_{x_{1}\sim p(x_{1}),x_{2}\sim p_{\varphi_{1}}(x_{2}|x_{1})}[log q_{\varphi_{2}}(x_{1}|x_{2})]= L^{x_{1}\rightarrow \hat{x}_{1}}(\varphi_{1},\varphi_{2})
\end{split}
\end{equation}
Similarly,
\begin{equation}
\begin{split}
&H(x_{2}|x_{1})\\
=&-E_{x_{2}\sim q(x_{2}),x_{1}\sim q_{\varphi_{2}}(x_{1}|x_{2})}[log q_{\varphi_{2}}(x_{2}|x_{1})]\\
\leq &-E_{x_{2}\sim q(x_{2}),x_{1}\sim q_{\varphi_{2}}(x_{1}|x_{2})}[log p_{\varphi_{1}}(x_{2}|x_{1})]= L^{x_{2}\rightarrow \hat{x}_{2}}(\varphi_{1},\varphi_{2})
\end{split}
\end{equation}
where the $\hat{x}_{1}$ and $\hat{x}_{2}$ are denoted as the reconstructions of $x_{1}$ and $x_{2}$. KL denotes the Kullback-Leible divergence. According to the equations of  $(2)$ and $(3)$,  on the one hand, we have a function $G_{3}:x_{1}\rightarrow \hat{x}_{1}$ defined by $G_{3}=G_{1} \circ G_{2}$, which first generates $x_{2}$ from $x_{1}$ based on $G_{1}$, then $G_{2}$ produces $\hat{x}_{1}$ from generated $x_{2}$. On the other hand, we also have a function $G_{4}:x_{2}\rightarrow \hat{x}_{2}$ defined by $G_{4}=G_{2} \circ G_{1}$, which first generates $x_{1}$ from $x_{2}$ based on $G_{2}$, then $G_{1}$ produces $\hat{x}_{2}$ from generated $x_{1}$. In contrast  to  the  fully adversarial training for solving $L^{x_{1}\rightarrow \hat{x}_{1}}(\varphi_{1},\varphi_{2})$ and $L^{x_{2}\rightarrow \hat{x}_{2}}(\varphi_{1},\varphi_{2})$, we employ the reconstruction loss to reduce the difficulty of model training. Specifically, we consider following object:
\begin{equation}
\begin{split}
\mathop{\min_{\varphi_{1},\varphi_{2}}} &O^{x_{1}\rightarrow \hat{x}_{1}}(\varphi_{1},\varphi_{2})\\
&=E_{\hat{x}_{1}\sim q_{\varphi_{2}}(\hat{x}_{1}|x_{2}),x_{2}\sim p_{\varphi_{1}}(x_{2}|x_{1}) }\ L_{mae}(x_{1},\hat{x}_{1})
\end{split}
\end{equation}
\begin{equation}
\begin{split}
\mathop{\min_{\varphi_{1},\varphi_{2}}} &O^{x_{2}\rightarrow \hat{x}_{2}}(\varphi_{1},\varphi_{2})\\
&=E_{\hat{x}_{2}\sim p_{\varphi_{1}}(\hat{x}_{2}|x_{1}),x_{1}\sim q_{\varphi_{2}}(x_{1}|x_{2})}\ L_{mae}(x_{2},\hat{x}_{2})
\end{split}
\end{equation}
where the $L_{mae}(\cdot)$ denotes the mean absolute error. Finally, we have the following object for BAI:
\begin{equation}
\begin{split}
\mathop{\min_{\varphi_{1},\varphi_{2}}\max_{\psi_{1}}}\ &\lambda_{d}O^{d}(\varphi_{1},\varphi_{2},\psi_{1}) \\
&+\lambda_{r}O^{x_{1}\rightarrow \hat{x}_{1}}(\varphi_{1},\varphi_{2})\\
&+\lambda_{r}O^{x_{2}\rightarrow \hat{x}_{2}}(\varphi_{1},\varphi_{2})
\end{split}
\end{equation}
where $\lambda_{d}$ and $\lambda_{r}$ are hyperparameters to balance the adversarial loss and the reconstruction loss.

\begin{figure}[!hbtp]
\begin{center}
\scalebox{.99}{
   \includegraphics[width=1\linewidth]{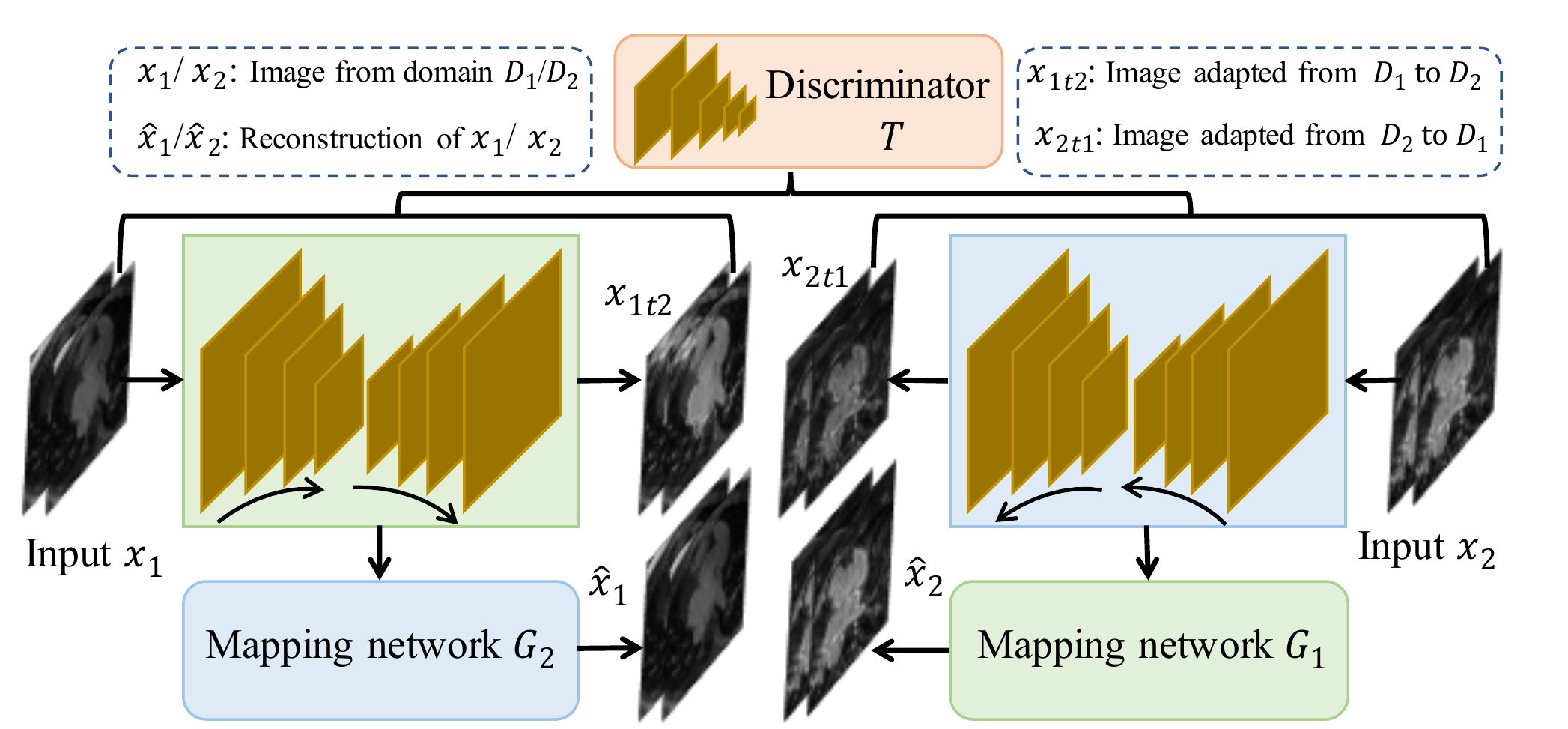}
   }
\end{center}
\caption{Structure of bidirectional adversarial inference network. The mapping network $G_{1}$ and the mapping network $G_{2}$ have the same structure.}
\label{fig:BAI} 
\end{figure}

\subsection{Hierarchical Dual Consistency for Semi-supervised Segmentation} 
The BAI makes the cross-domain data adapt to each other to produce matched domains.  In detail, the domain $D_{1}$ adapted to source $D_{2}$ is denoted as $D_{1t2}=D^{l}_{1t2} \cup D^{u}_{1t2}$, where $D^{l}_{1t2}=\{((x^{l}_{1t2})^{i},y^{i})\}^{n_{1}}_{i=1}$ with $n_{1}$ labelled samples and $D^{u}_{1t2}=\{(x^{u}_{1t2})^{i}\}^{n_{1}+n_{2}}_{i=n_{1}+1}$ with $n_{2}$ unlabelled samples. The domain $D_{2}$ adapted to source $D_{1}$ is denoted as $D_{2t1}=D^{u}_{2t1}=\{(x^{u}_{2t1})^{i}\}^{m}_{i=1}$ with $m$ unlabelled samples. Then we merge two source domains and two adapted domains to obtain the matched domains of $D_{p1}$ and $D_{p2}$. The $D_{p1}=D_{1}\cup D_{2t1}=D^{l}_{1}\cup D^{u}_{1}\cup D^{u}_{2t1}=D^{l}_{p1} \cup D^{u}_{p1}$, where $D^{l}_{p1}=\{((x^{l}_{p1})^{i},y^{i})\}^{n_{1}}_{i=1}$ with $n_{1}$ labelled samples and $D^{u}_{p1}=\{(x^{u}_{p1})^{i}\}^{n_{1}+n_{2}+m}_{i=n_{1}+1}$ with $n_{2}+m$ unlabelled samples. The $D_{p2}=D_{2}\cup D_{1t2}=D^{l}_{1t2}\cup D^{u}_{1t2}\cup D^{u}_{2}=D^{l}_{p2} \cup D^{u}_{p2}$, where $D^{l}_{p2}=\{((x^{l}_{p2})^{i},y^{i})\}^{n_{1}}_{i=1}$ with $n_{1}$ labelled samples and $D^{u}_{p2}=\{(x^{u}_{p2})^{i}\}^{n_{1}+n_{2}+m}_{i=n_{1}+1}$ with $n_{2}+m$ unlabelled samples. We denote two domain marginal distributions of $D_{p1}$ and $D_{p2}$ as $p(x_{p1})$ and $q(x_{p2})$, respectively. The joint distribution of $D_{p1}$ and $D_{p2}$ is denoted as $j(x_{p1}, x_{p2})$.

Based on the matched domains, we investigate complementary LA modelling and complementary domain knowledge learning to provide inherent prediction perturbation for the consistency based cross-domain semi-supervised learning. Therefore, a hierarchical dual consistency is investigated. Specifically, for the intra-domain, we consider two dual-modelling networks $S_{1}:x_{p1}\rightarrow (\widetilde{y}_{l1},\widetilde{y}_{g1})$ parameterised by $\theta_{1}=\{\theta^{f}_{1},\theta^{l}_{1},\theta^{g}_{1}\}$ and $S_{2}:x_{p2}\rightarrow (\widetilde{y}_{l2},\widetilde{y}_{g2})$ parameterised by $\theta_{2}=\{\theta^{f}_{2},\theta^{l}_{2},\theta^{g}_{2}\}$ applied to the matched domains of $D_{p1}$ and $D_{p2}$, respectively. Each dual-modelling network estimates two targets by considering local information and global information of image, where $S_{1}$ simultaneously performs the global modelling of $S_{g1}:x_{p1}\rightarrow \widetilde{y}_{g1}$ parameterised by $\{\theta^{f}_{1},\theta^{g}_{1}\}$ and the local modelling of $S_{l1}:x_{p1}\rightarrow \widetilde{y}_{l1}$ parameterised by $\{\theta^{f}_{1},\theta^{l}_{1}\}$. Similarly, the $S_{2}$ simultaneously performs the global modelling of $S_{g2}:x_{p2}\rightarrow \widetilde{y}_{g2}$ parameterised by $\{\theta^{f}_{2},\theta^{g}_{2}\}$ and the local modelling of $S_{l2}:x_{p2}\rightarrow \widetilde{y}_{l2}$ parameterised by $\{\theta^{f}_{2},\theta^{l}_{2}\}$. Then we encourage the global modelling and the local modelling of each dual-modelling network to predict consistent targets via the consistency loss:
\begin{equation}
\begin{split}
\min_{\theta_{1}}\ O^{intra1}(\theta_{1})=E_{x^{u}_{p1} \sim p(x_{p1})}L_{d}(S_{l1}(x^{u}_{p1}),S_{g1}(x^{u}_{p1}))
\end{split}
\end{equation}
\begin{equation}
\begin{split}
\min_{\theta_{2}}\ O^{intra2}(\theta_{2})=E_{x^{u}_{p2} \sim q(x_{p2})}L_{d}(S_{l2}(x^{u}_{p2}),S_{g2}(x^{u}_{p2}))
\end{split}
\end{equation}
where $L_{d}(\cdot)$ denotes the dice loss function. For the dual consistency in inter-domain,  we maximise the agreement on two matched domains. Therefore, we encourage $S_{1}$ and $S_{2}$ to predict similar outputs by:
\begin{equation}
\begin{split}
\min_{\theta_{1},\theta_{2}}\ &O^{inter}(\theta_{1},\theta_{2})\\
=&E_{(x^{u}_{p1},x^{u}_{p2}) \sim j(x_{p1},x_{p2})} L_{c}(S_{1}(x^{u}_{p1}),S_{2}(x^{u}_{p2}))\\
=&E_{(x^{u}_{p1},x^{u}_{p2}) \sim j(x_{p1},x_{p2})}(L_{c}(S_{l1}(x^{u}_{p1}),S_{l2}(x^{u}_{p2}))\\
+&L_{c}(S_{g1}(x^{u}_{p1}),S_{g2}(x^{u}_{p2})))
\end{split}
\end{equation}
where $L_{c}(\cdot)$ denotes the cross-entropy loss function. To avoid that $S_{1}$ and $S_{2}$ gradually resemble each other, we encourage the $S_{1}$ and $S_{2}$ to produce conditional independent features by orthogonalising the weights of feature layers: 
\begin{equation}
\begin{split}
\min_{\theta_{1},\theta_{2}}\ O^{ow}(\theta_{1},\theta_{2})=\frac{1}{N}\sum^{N}_{i=1}(\frac{1}{K^{2}_{i}}\sum^{K^{2}_{i}}|\frac{(\theta^{f}_{1i})^{T}\theta^{f}_{2i}}{||\theta^{f}_{1i}||||\theta^{f}_{2i}||}|)
\end{split}
\end{equation}
where the $N$ denotes the number of layers in $S_{1}$ and $S_{2}$. $K_{i}$ represents the number of features in $i$th layer. $\theta^{f}_{1i}$ and $\theta^{f}_{2i}$ denote the parameters of $i$th feature layer in $S_{1}$ and $S_{2}$, respectively. 

Beyond the consistency learning above, $S_{1}$ and $S_{2}$ can explicitly learns from $D^{l}_{p1}$ and $D^{l}_{p2}$ with the supervision of the labels:
\begin{equation}
\begin{split}
\min_{\theta_{1}}\ O^{super1}(\theta_{1})=&E_{x^{l}_{p1} \sim p(x_{p1})} L_{s}(S_{1}(x^{l}_{p1}),y)\\
=&E_{x^{l}_{p1} \sim p(x_{p1})}(L_{s}(S_{l1}(x^{l}_{p1}),y)\\
+&L_{s}(S_{g1}(x^{l}_{p1}),y))
\end{split}
\end{equation}
\begin{equation}
\begin{split}
\min_{\theta_{2}}\ O^{super2}(\theta_{2})=&E_{x^{l}_{p2} \sim q(x_{p2})}L_{s}(S_{2}(x^{l}_{p2}),y)\\
=&E_{x^{l}_{p2} \sim q(x_{p2})}L_{s}(S_{l2}(x^{l}_{p2}),y)\\
+&L_{s}(S_{g2}(x^{l}_{p2}),y))
\end{split}
\end{equation}
where the $y$ denotes the LA label. $L_{s}(\cdot)$ denotes the supervised loss functions (cross-entropy loss function and dice loss function). Then the final training objective for the learning of $S_{1}$ and $S_{2}$ is denoted as:
\begin{equation}
\begin{split}
\min_{\theta_{1},\theta_{2}}\ O^{total}(\theta_{1},\theta_{2})&=\lambda_{super}(O^{super1} +O^{super2})\\
&+\lambda_{intra}(O^{intra1}+O^{intra2})\\
&+\lambda_{inter}O^{inter}+\lambda_{ow}O^{ow}\\
\end{split}
\end{equation}
where the $\lambda_{super}$, $\lambda_{intra}$, $\lambda_{inter}$ and $\lambda_{ow}$ are hyperparameters to balance the loss terms.

\begin{figure}[!hbtp]
\begin{center}
\scalebox{.99}{
   \includegraphics[width=1\linewidth]{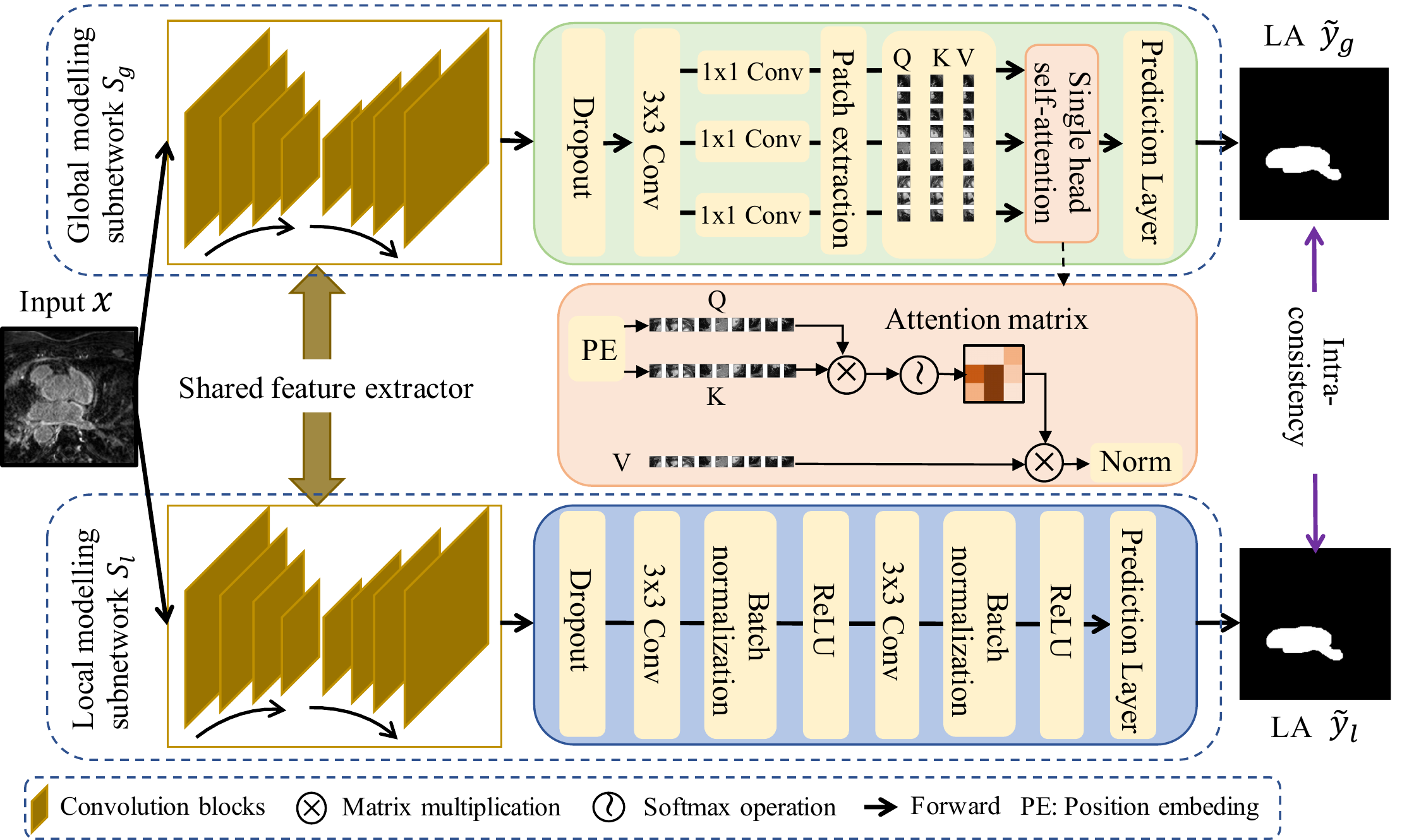}
   }
\end{center}
\caption{Dual-modelling network for intra-consistency learning. The local-modelling branch and global-modelling branch share a feature extractor. For the global-modelling branch, the extracted feature maps from input images are split into $8\times 8$ patches. These $8\times 8$ patches are taken as a sequence of vectors to be fed to a self-attention based global-modelling structure.}
\label{fig:global_modelling} 
\end{figure}

\subsection{Network Configuration} 
The BAI module contains three subnetworks: two domain mapping networks ($G_{1}$, $G_{2}$) and a discriminative network $T$. We use the 2D U-Net with bilinear upsampling as network backbones of both $G_{1}$ and $G_{2}$. $T$ has six convolution layers with the numbers of filters of $32, 64, 128, 256, 256, 1$, respectively. Each of the first five $3\times 3$ convolutional layers with a stride of $2$ is followed by a batch normalisation layer and a ReLU layer. The final $1\times 1$ convolutional layer with a stride of $1$ is followed by a sigmoid layer.

Hierarchical dual-modelling network contains two dual-modelling networks with the same structure. Each dual-modelling network contains a 2D U-Net with bilinear upsampling used to extract image features and two branch networks used to estimate targets. The two branch networks are the global modelling network and the local modelling network. The global modelling network is based on the self-attention \cite{dosovitskiy2021image,carion2020end,wang2018non} as shown in Fig. \ref{fig:global_modelling}. In the global modelling network, we use the sinusoidal position encoding to emphasise the sequential relationship between input feature patches \cite{vaswani2017attention}. The local modelling network consists of three convolution blocks. The details are shown in Fig. \ref{fig:global_modelling}.

\begin{table*}
\captionsetup{justification=centering}
 \setlength{\abovecaptionskip}{0pt} 
 \setlength{\belowcaptionskip}{0pt} 
 \caption{Comparison of four LGE-CMRI datasets from different centres.  Abbreviations: TE, Echo Time; TR, Repetition Time; CARMA, Comprehensive Arrhythmia Research and Management.}
 \centering
 \scalebox{.9}{
 \setlength{\floatsep}{10pt plus 3pt minus 2pt}
\begin{tabular}{cccccc}
\addlinespace
\toprule
Centres & \multicolumn{1}{c}{Acquired Resolution} & \multicolumn{1}{c}{TE/TR} & \multicolumn{1}{c}{Scanner} & \multicolumn{1}{c}{Source} &\multicolumn{1}{c}{Amount of Data}\\ \midrule
 C1 & $(1.4\sim 1.5)\times (1.4\sim 1.5)\times 4\ $mm$^{3}$    &2.2$/$5.2 ms       &1.5 Tesla Avanto  &Royal Brompton Hospital   & $165\ $ LGE-CMR scans \\ \midrule
 C2 & $1.25\times 1.25\times 2.5\ $  mm$^{3}$                  &2.3/5.4 ms        &\tabincell{c}{1.5 Tesla Avanto,\\ 3.0 Tesla Vario}  & CARMA, University of Utah & $153\ $ LGE-CMR scans \\ \midrule
 C3 &$1.4\times 1.4\times 1.4\ $mm$^{3}$                 &2.1/5.3 ms  &1.5T Philips Achieva  &Beth Israel Deaconess Medical Center     & $20\ $ LGE-CMR scans \\ \midrule
  C4 &$1.3\times 1.3\times 4\ $mm$^{3}$               &2.1/5.3 ms  &1.5T Philips Achieva  &Imaging Sciences at King’s College London     & $20\ $ LGE-CMR scans \\ \midrule
\end{tabular}
 }
 \label{table:datasets} 
\end{table*}

\section{Experiments}
\subsection{Overview of Experiments}
Comprehensive experiments were performed to validate our proposed AHDC.

\textbf{(1) The feasibility of AHDC for generalising across domains}: Our proposed AHDC was validated on four 3D late gadolinium enhancement cardiac MR (LGE CMR) datasets and a 3D CT dataset combined in pairs, which followed the independent validation protocol. Furthermore, we also investigated the impact of different ratios ($r=\{5\%, 10\%, 20\%\}$) of the labelled data for validating our proposed AHDC.

\textbf{(2) The superiority of AHDC for generalising across domains}: We compared to widely used and state-of-the-art semi-supervised methods on cross-domain data for comparison, including mean teacher (MT) method \cite{tarvainen2017mean}, uncertainty-aware self-ensembling model (UA-MT) \cite{yu2019uncertainty}, Dual-Task consistency (DTC) \cite{luo2021semi} and Dual-Teacher \cite{li2020dual}. It is of note that MT, UA-MT and DTC were proposed for the single-domain semi-supervised learning while the Dual-Teacher method was proposed for the cross-domain learning. Besides, the Dual-Teacher required the labelled data from both cross-domain data for model learning. For a fair comparison, MT, UA-MT and DTC were performed on one of the matched domains, i.e., $D_{p1}$. We also compared with the joint training method that combining the cross-domain data directly for the LA segmentation based on our proposed semi-supervised method.

\textbf{(3) The effectiveness of the components in AHDC}: Firstly, we compared the performance between different architectures of the BAI module. On the one hand, to validate the effectiveness of bidirectional reconstruction for specifying the relationship of matched samples, an experiment was performed on bidirectional adversarial inference without using bidirectional reconstruction (BAI$_{wbr}$/ALI/BiGAN). On the other hand, to validate the effectiveness of skip connection of domain mapping network for keeping target structure consistent, an experiment was performed on bidirectional adversarial inference without using skip connection in domain mapping network (BAI$_{eds}$). Then, we further validated the performance of BAI by comparing it with the fully adversarial ALICE \cite{li2017alice} on the downstream semi-supervised tasks. Finally, for validating the effectiveness of HDC,  we decomposed the HDC into independent intra-domain dual consistency learning (HDC$_{intra}$) by removing a dual-modelling network and inter-domain dual consistency learning (HDC$_{inter}$) by removing global modelling branch but retaining local modelling branch.

\textbf{(4) The effectiveness of the BAI for matching domains}:  Firstly, we performed the principal components analysis to show the data distributions of source domains ($D_{1}$ and $D_{2}$) and the adapted domains ($D_{1t2}$ and $D_{2t1}$). The data distributions of source domains and the adapted domains were compared to validate the effectiveness of AHDC for aligning distributions. Then,  we made a qualitative visualisation of images before and after the bidirectional adversarial inference to validate the effectiveness of AHDC for matching samples.

\textbf{(5) The effectiveness of the HDC for the availability of complementary information}: To validate the availability of complementary modelling information in the intra-domain, we compared the segmentation performance of dual modelling network (local-global modelling structure) to the ones without using dual-modelling structures. Specifically, we replaced the local-modelling branch with the global-modelling branch (global-global modelling structure) and replaced the global-modelling branch with the local-modelling branch (local-local modelling structure) in dual modelling network for experiments. To validate the availability of complementary domain information in inter-domain, we compared the segmentation performance of HDC with/without using the orthogonal weight constraint (WOW and WOOW).

\textbf{(6) The effects of parameter settings on model performance}: We explored two important parameter settings. (\romannumeral1) The impact of different patch sizes ($4\times 4$, $8\times 8$ and $16\times 16$) for global modelling. (\romannumeral2) The impact of different values of $\lambda_{ow}$ (0.0, 0.1, and 1.0 ) for inter-domain learning.

\subsection{Datasets} 
To evaluate the performance of our proposed AHDC, four 3D LGE-MRI datasets (C1, C2, C3 and C4) and a 3D CT dataset (C5) were collected as a retrospective study. In our experiments, the collected datasets of C1 and C2 included segmentation of the LA epicardium and LA endocardium while the collected datasets of C3, C4 and C5 included segmentation of the LA endocardium. We have summarised the characteristics of the four 3D LGE-MRI datasets to emphasise their differences as shown in TABLE \ref{table:datasets}. 
 
LGE-MRI scanning sequence of centre 1 (C1): Cardiac MR data were acquired in patients with longstanding persistent atrial fibrillation (AF) on a Siemens Magnetom Avanto 1.5T scanner (Siemens Medical Systems, Erlangen, Germany). Transverse navigator-gated 3D LGE-CMRI \cite{peters2009recurrence} was performed using an inversion prepared segmented gradient echo sequence (TE/TR 2.2ms/5.2ms) 15 minutes after gadolinium administration (Gadovist-gadobutrol, 0.1mmol/kg body weight, BayerSchering, Berlin, Germany) \cite{haissaguerre1998spontaneous}. The inversion time was set to null the signal from normal myocardium. The acquired resolution parameter of LGE-CMRI data was $(1.4-1.5)\times (1.4-1.5)\times 4$ mm$^{3}$ (reconstructed to $(0.7-0.75)\times (0.7-0.75)\times 2$ mm$^{3}$). LGE-CMRI data were acquired during free-breathing using a crossed-pairs navigator positioned over the dome of the right hemi-diaphragm with navigator acceptance window size of $5mm$ and CLAWS respiratory motion control \cite{keegan2014improved,keegan2014navigator}. The LGE CMR data were collected from the Royal Brompton Hospital. In total, 165 scans were used in this study. 

LGE-MRI scanning sequence of centre 2 (C2): Cardiac MR data were obtained on a 1.5 Tesla Avanto scanners or a 3.0 Tesla Vario (Siemens Medical Solutions, Erlangen, Germany). The scan is acquired 20–25 minutes after 0.1 mmol/kg gadolinium contrast  (Multihance, Bracco Diagnostics Inc., Princeton, NJ) using a 3D respiratory navigated, inversion recovery prepared gradient echo pulse sequence. Typical acquisition parameters are free breathing using navigator gating, a transverse imaging volume with voxel size = $1.25\times 1.25\times 2.5$ mm$^{3}$ (reconstructed to $0.625\times 0.625\times 2.5$ mm$^{3}$), TR/TE = 5.4/2.3 ms, inversion time (TI)=270-310 ms. The TI value for the LGE-MRI scan is identified using a scout scan. Typical scan times for the LGE-MRI study were between 8 and 15 min at 1.5 T and 6–11 min using the 3T scanner (for Siemens sequences) depending on subject respiration and heart rates. The LGE CMR data were collected from the Comprehensive Arrhythmia Research and Management, University of Utah. In total, 153 scans were used in this study.

LGE-MRI scanning sequence of center 3 (C3): C3 is from the ISBI 2012 Left Atrium Fibrosis and Scar Segmentation Challenge \cite{karim2013evaluation,li2021atrialgeneral}. The LGE CMR data were collected from the Beth Israel Deaconess Medical Center. In total, 20 scans were used in this study. 

LGE-MRI scanning sequence of center 4 (C4): C4 is also from the ISBI 2012 Left Atrium Fibrosis and Scar Segmentation Challenge \cite{karim2013evaluation,li2021atrialgeneral}. The LGE CMR data were collected from the Imaging Sciences at King’s College. In total, 20 scans were used in this study. 

CT scanning sequence of centre 5 (C5): C5 is from the Multi-modality Whole Heart Segmentation (MM-WHS) 2017 dataset \cite{zhuang2016multi,zhuang2013challenges,zhuang2015multiatlas,zhuang2010registration}. In total, 60 CT scans were used in this study.

\subsection{Experimental Setup}
\textbf{(1) Data partitioning}: For C1, the 3D LGE-MRI dataset with 165 scans was randomly split into a training set with 99 scans and a testing set with 66 scans (33 pre-ablation scans and 33 post-ablation scans). The training set then was randomly split into a labelled training set with 20 scans (20\%) and an unlabelled training set with 79 scans (80\%). For C2, the 3D LGE-MRI dataset with 153 scans was randomly split into a training set with 91 scans and a testing set with 62 scans (31 pre-ablation scans and 31 post-ablation scans). The training set then was randomly split into a labelled training set with 18 scans (20\%) and an unlabelled training set with 73 scans (80\%).  For C3 and C4, each 3D LGE-MRI dataset with 20 scans was randomly split into a training set with 12 scans and a testing set with 8 scans (4 pre-ablation scans and 4 post-ablation scans). The training set then was randomly split into a labelled training set with 4 scans and an unlabelled training set with 8 scans. Because C5 only provides 60 CT scans including 20 labelled scans and 40 unlabelled scans, we randomly selected 15 scans from 20 labelled scans as a testing set. The remaining 5 labelled scans (labelled training set) and 40 unlabelled scans (unlabelled training set) together as a training set. Since each patient may contain multiple 3D LGE-MRI scans, the 3D LGE-MRI datasets  were split under the strategy that all scans from each unique patient were only in one of the training or testing sets. 

\textbf{(2) Implementation details}: Experiments were performed on five datasets combined in pairs for cross-centre study (C1 and C2, C3 and C4) and cross-modality study (C2 and C5). To reduce the dependence of models on annotated data and to avoid the impact of label variations from  different centres, there were two kinds of experiment settings for each cross-domain data. Take experiments on C1 and C2 as an example: one used C1 to support C2 that the model was trained using the labelled training set (18 labelled cases) of C2, the unlabelled training set (73 unlabelled cases) of C2 and the whole training set (99 unlabelled cases) of C1. The other one used C2 to support C1 that the model was trained using the labelled training set (20 labelled cases) of C1, the unlabelled training set (79 unlabelled cases) of C1 and the whole training set (91 unlabelled cases) of C2. We denoted the results obtained by the fully supervised model trained with the labelled training set from C1 (20 cases), C2 (18 cases), C3 (4 cases), C4 (4 cases) and C5 (5 cases) as the baseline and the results obtained by the fully supervised model trained with the whole training set from C1 (99 cases), C2 (91 cases), C3 (12 cases)  and C4 (12 cases) as the upper bound. 

We pre-processed the data with the normalisation. Smaller patches of $256\times 256$ centred on the LA region were cropped. To avoid overfitting, we applied data augmentations with random rotation. The training time of our model is about 17.17 hours while the testing time for one 3D case is about 0.259 seconds. For the learning of the BAI network, we used the Adam method to perform the optimisation of two mapping networks with an initial learning rate of $0.001$ and a decayed rate of $0.98$. The optimiser used in the discriminative network was Adam with a fixed learning rate of $0.0001$. For the learning of two dual-modelling networks, we also used the Adam method with an initial learning rate of $0.001$ and a decayed rate of $0.98$. The current statistics of batch normalisation were used for both training and testing. All experiments were performed with an independent test. For the dual consistency learning, in each iteration, we first performed the intra-consistency with both labelled and unlabelled data simultaneously, then performed the inter-consistency with both labelled and unlabelled data simultaneously,  performed supervised learning with labelled data in the last.  Our deep learning model was implemented using Tensorflow $1.2.1$ on an Ubuntu $16.04$ machine (The code will be released publicly once the manuscript is accepted for publication via https://github.com/Heye-SYSU/AHDC). It was trained and tested using an Nvidia RTX 8000 GPU (48GB GPU memory).

The coefficients $\lambda_{d}$ and $\lambda_{r}$ used to balance the adversarial loss and the reconstruction loss, were automatically learned based on the strategy of uncertainty \cite{kendall2018multi}. The coefficient $\lambda_{intra}$ was dynamically changed over time with the function of $f(t)=e^{-5*(1-\frac{t}{t_{max}})^{2}}$.  The coefficients $\lambda_{inter}$, $\lambda_{super}$ and $\lambda_{ow}$ were set to the values of $1.0$, $0.5$ and $0.1$, respectively.

\textbf{(3)  Evaluation criteria}: To evaluate the segmentation performance, we used region-based metrics \cite{dice1945measures,taha2015metrics}, e.g., the Dice Similarity Coefficient (DSC) and the Jaccard Index (JI), to validate the predicted segmentation map against the manually defined ground-truth. We also used a surface-based metric called Average Surface Distance (ASD) to provide the distance in $\mathrm{mm}$ to quantify the accuracy of the predicted mesh ($S$) compared to the ground-truth mesh ($S^\prime$)  \cite{taha2015metrics}. 

\begin{table*}
    \captionsetup{justification=centering}
    \caption{\centering Quantitative comparison between our proposed AHDC and other methods on multi-centre data. Abbreviations: DSC, Dice Similarity  Coefficient;  JI, Jaccard Index; ASD, Average Surface Distance.}\label{table:c1c2c3}\label{fig:test}
    \begin{subtable}[h]{0.99\textwidth}
        \caption{Experiments on C1 (MR) and C2 (MR)}
        \centering
            \begin{tabular}{c|ccc|ccc}
            \addlinespace
            \toprule
            \multirow{2}*{Method} & \multicolumn{3}{c|}{C2 (MR) supports C1 (MR)}  & \multicolumn{3}{c}{C1 (MR) supports C2 (MR)}\\ \cline{2-7}
                                 & DSC & JI  &ASD (mm)     & DSC & JI  &ASD (mm)\\ \hline
                                 
            Upper Bound     & $0.932\pm0.026$    & $0.874\pm0.044$    & $1.28\pm0.847$   & $0.926\pm0.021$    & $0.863\pm0.036$    & $0.867\pm0.46$   \cr 
            Baseline        & $0.869\pm0.078$    & $0.775\pm0.111$    & $2.81\pm2.08$    & $0.860\pm0.103$    & $0.765\pm0.122$    & $4.00\pm5.08$   \\ \hline
              MT            &$0.882\pm0.059$     &$0.793\pm0.090$ &$2.45\pm1.67$         &$0.880\pm0.071$     &$0.792\pm0.096$      &$1.79\pm1.96$    \cr
              UA-MT         &$0.885\pm0.060$     &$0.799\pm0.092$      & $2.19\pm1.47$   &$0.884\pm0.072$     &$0.799\pm0.098$      & $2.79\pm3.76$   \cr 
              DTC           &$0.887\pm0.061$     &$0.803\pm0.094$     &$2.25\pm1.54$     &$0.888\pm0.076$     &$0.806\pm0.102$    &$2.40\pm3.48$    \cr 
            Dual-Teacher    &$0.899\pm0.046$     &$0.820\pm0.073$     &$1.83\pm1.02$     &$0.896\pm0.064$     &$0.816\pm0.088$     &$1.97\pm3.13$   \cr \hline 
            Joint-training &$0.889\pm0.059$     &$0.805\pm0.091$     &$2.05\pm1.22$    &$0.887\pm0.056$     &$0.801\pm0.081$     &$1.59\pm1.56$  \cr \hline
              AHDC          & $\mathbf{0.916\pm0.041}$  &$\mathbf{0.848\pm0.066}$ & $\mathbf{1.47\pm0.846}$   & $\mathbf{0.917\pm0.026}$    &$\mathbf{0.848\pm0.043}$     & $\mathbf{1.17\pm1.60}$  \\ \midrule 
            \end{tabular}
    \end{subtable}
\vspace{0.25cm}
\\
    \begin{subtable}[h]{0.99\textwidth}
        \caption{Experiments on C3 (MR) and C4 (MR)}
        \centering
            \begin{tabular}{c|ccc|ccc}
                \addlinespace
                \toprule
                \multirow{2}*{Method} & \multicolumn{3}{c|}{C4 (MR) supports C3 (MR)}  & \multicolumn{3}{c}{C3 (MR) supports C4 (MR)}\\ \cline{2-7}
                                     & DSC & JI  &ASD (mm)      & DSC & JI  &ASD (mm)\\ \hline
                                     
                Upper Bound     & $0.808\pm0.035$    & $0.679\pm0.050$    & $2.42\pm0.645$   & $0.841\pm0.043$    & $0.727\pm0.062$    & $2.07\pm0.543$     \cr   
                Baseline         & $0.684\pm0.098$     & $0.528\pm0.109$    & $6.17\pm3.63$  & $0.742\pm0.081$    & $0.596\pm0.095$    & $4.20\pm1.08$       \\ \hline
                  MT             &$0.749\pm0.073$      &$0.604\pm0.092$ &$3.59\pm1.77$     &$0.797\pm0.101$    &$0.673\pm0.122$ &$2.42\pm1.10$               \cr
                  UA-MT           &$0.760\pm0.081$      &$0.619\pm0.100$ &$3.96\pm2.10$    &$0.811\pm0.086$    &$0.690\pm0.108$ &$2.21\pm0.718$            \cr 
                  DTC            &$0.765\pm0.066$       &$0.624\pm0.086$ &$3.40\pm1.72$    &$0.809\pm0.088$    &$0.687\pm0.111$ &$2.94\pm0.935$             \cr 
                Dual-Teacher     &$0.773\pm0.050$       &$0.633\pm0.067$ &$3.03\pm0.988$   &$0.817\pm0.089$    &$0.699\pm0.112$ &$2.48\pm0.937$             \cr \hline 
                Joint-training   &$0.770\pm0.056$     &$0.629\pm0.074$     &$3.45\pm1.51$  &$0.811\pm0.094$     &$0.691\pm0.115$     &$2.70\pm0.889$          \cr \hline
                  AHDC            & $\mathbf{0.795\pm0.044}$    & $\mathbf{0.661\pm0.061}$  &  $\mathbf{2.47\pm0.681}$  & $\mathbf{0.830\pm0.057}$    & $\mathbf{0.713\pm0.077}$  &  $\mathbf{2.07\pm0.659}$  \\ \midrule 
                \end{tabular}
    \end{subtable}
\end{table*}

\begin{table*}
 \captionsetup{justification=centering}
 \caption{\centering Quantitative comparison between our proposed AHDC and other methods on multi-modality data. Abbreviations: DSC, Dice Similarity  Coefficient;  JI, Jaccard Index; ASD, Average Surface Distance.}\label{table:MR_CT}
 \centering
\scalebox{.99}{
    \begin{tabular}{c|ccc|ccc}
        \addlinespace
        \toprule
        \multirow{2}*{Method} & \multicolumn{3}{c|}{C5 (CT) supports C2 (MR)}  & \multicolumn{3}{c}{C2 (MR) supports C5 (CT)}\\ \cline{2-7}
                              & DSC & JI  &ASD (mm)     & DSC & JI  &ASD (mm)\\ \hline
        Upper Bound    & $0.923\pm0.025$    & $0.858\pm0.042$    & $1.20\pm1.96$  & -    & -    & -  \cr               
        Baseline       & $0.858\pm0.107$    & $0.763\pm0.121$    & $2.72\pm4.59$   & $0.828\pm0.115$    & $0.722\pm0.157$    & $4.88\pm3.53$  \\ \hline
          MT            &$0.874\pm0.072$    &$0.782\pm0.99$ &$2.42\pm3.14$   &$0.861\pm0.086$    &$0.765\pm0.121$ &$4.54\pm3.35$    \cr
          UA-MT         &$0.881\pm0.056$    &$0.791\pm0.084$ &$1.84\pm2.66$   &$0.878\pm0.050$    &$0.786\pm0.077$ &$2.39\pm1.47$  \cr 
          DTC           &$0.888\pm0.059$    &$0.803\pm0.084$ &$1.93\pm3.29$   &$0.880\pm0.064$    &$0.791\pm0.096$ &$2.96\pm2.10$   \cr 
        Dual-Teacher    &$0.888\pm0.041$    &$0.801\pm0.062$ &$1.41\pm1.47$   &$0.891\pm0.036$      &$0.806\pm0.057$ &$2.38\pm1.14$   \cr \hline
        Joint-training &$0.869\pm0.069$     &$0.774\pm0.100$     &$2.18\pm3.89$    &$0.834\pm0.117$     &$0.731\pm0.156$     &$5.62\pm5.10$  \cr \hline
          AHDC           & $\mathbf{0.911\pm0.028}$    & $\mathbf{0.837\pm0.047}$  &  $\mathbf{1.07\pm0.872}$   & $\mathbf{0.916\pm0.031}$    & $\mathbf{0.846\pm0.052}$  &  $\mathbf{1.30\pm0.319}$  \\ \midrule 
        \end{tabular}
        }
\end{table*}

\section{Results and Analysis}
In this section, we demonstrate the results of the above mentioned experiments to validate our proposed AHDC for the cross-domain semi-supervised segmentation.

\subsection{The Feasibility Analysis of AHDC for Generalising Across Domains:}
\begin{figure}[!hbtp]
\begin{center}
\scalebox{.99}{
   \includegraphics[width=1\linewidth]{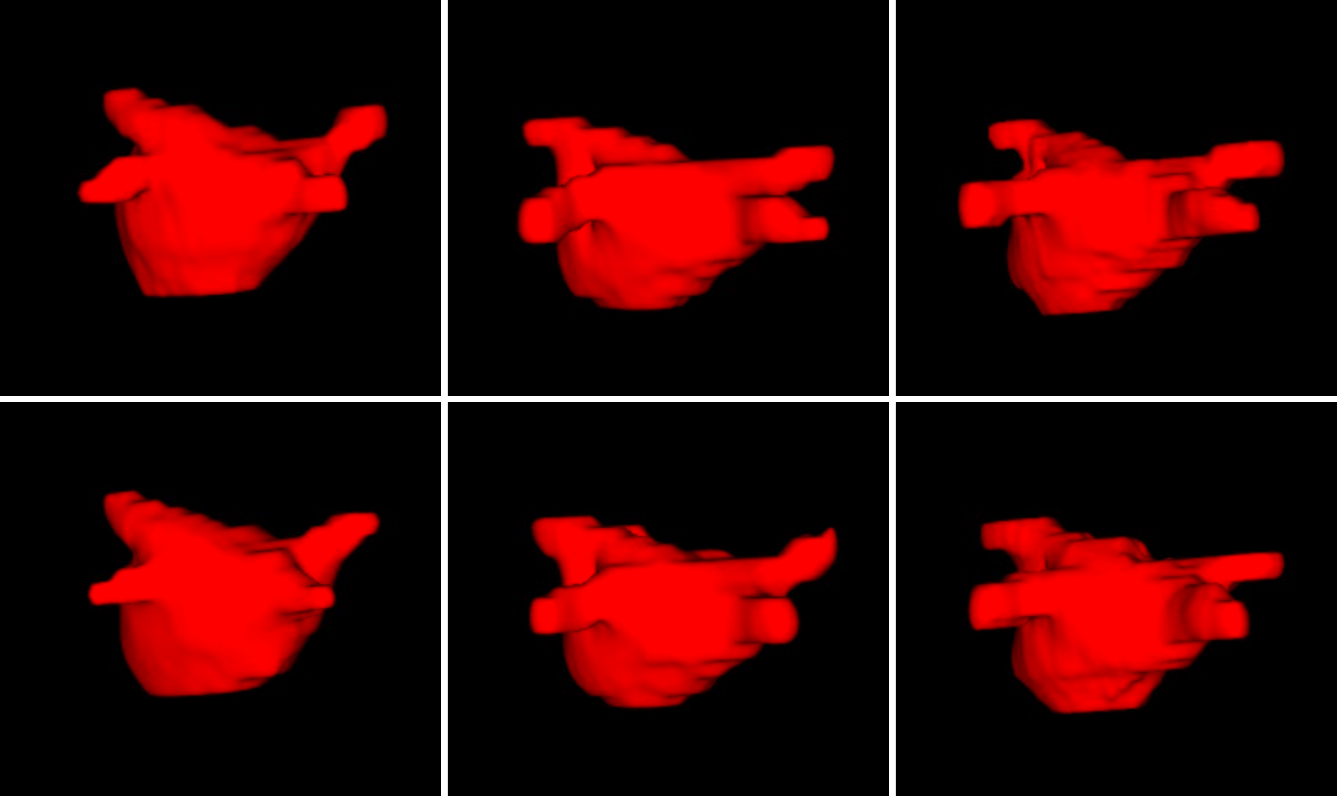}
   }
\end{center}
\caption{3D visualization of LA segmentation results estimated by AHDC. Each DSC score is calculated for the whole 3D LGE-MRI image (The DSC based segmentation accuracies of AHDC for the 3D slices from column $1$ to column $3$ are  $0.936$, $0.917$, and $0.898$, respectively). Abbreviations: DSC, Dice Similarity  Coefficient.}
\label{fig:vis_3d} 
\end{figure}

TABLE \ref{table:c1c2c3} and TABLE \ref{table:MR_CT} summarises the quantitative segmentation results of AHDC on multi-centre data and multi-modality data. As we can see, our proposed AHDC obtains consistent improvements in terms of the DSC, JI and ASD against the baselines. Furthermore, as the experiment results are summarised in TABLE \ref{table:ratio_label}, one can see that our proposed AHDC obtains consistent improvements against the fully supervised learning under the $5\%$, $10\%$, $20\%$ labelled data setting. Fig. \ref{fig:la_vis_comparison} and Fig. \ref{fig:vis_3d} provide the 2D and 3D qualitative LAs estimated by AHDC compared to the ground truth. It is observed that our proposed AHDC has the ability to segment LA accurately. These quantitative and qualitative results indicate the feasibility of our proposed AHDC for generalising across domains.

\subsection{The Superiority Analysis of AHDC for Generalising Across Domains:}
\begin{figure*}[!ht]
	\begin{center}
	\scalebox{.85}{
	 \includegraphics[width=1\textwidth]{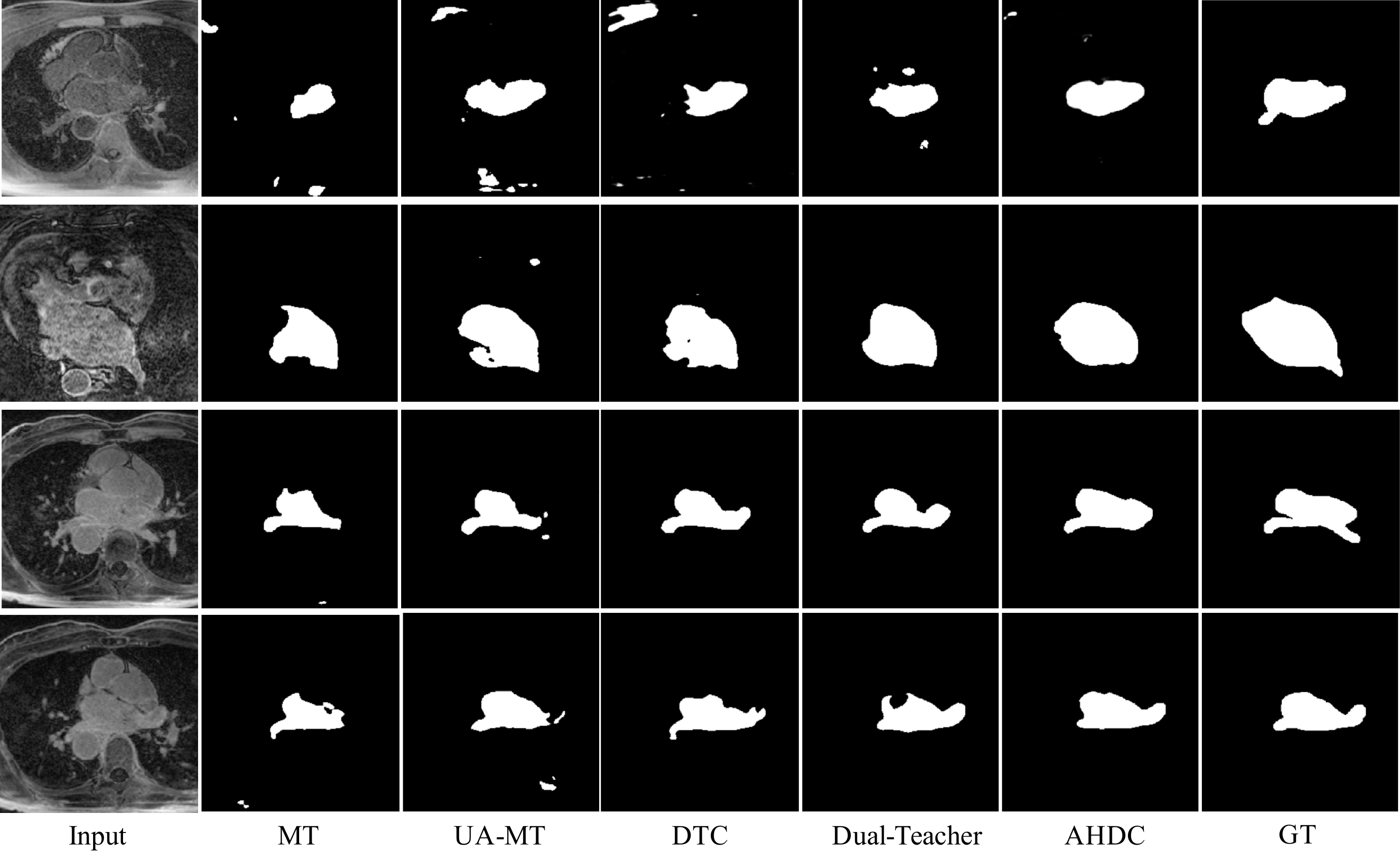}
	 }
	\end{center}
	\caption{2D visual comparisons on LA segmentation results estimated by different methods. It is observed that our estimated LAs (AHDC) are more similar to the ground truth (GT) than others (DSC based segmentation accuracies of AHDC for the 2D slices from row 1 to row 4 are  $0.859$, $0.897$, $0.907$ and $0.949$, respectively). Abbreviations: DSC, Dice Similarity  Coefficient.}\label{fig:la_vis_comparison}
\end{figure*}

TABLE \ref{table:c1c2c3} and TABLE \ref{table:MR_CT}  summarises the experiment results on multi-centre data and multi-modality data combined in pairs for comparison. It is observed that the widely used semi-supervised method of MT improves the segmentation accuracy of LA compared to the baseline. One can see that after adding uncertainty information to the MT, the performance of the MT is improved (UA-MT). The DTC method further improves the segmentation accuracy, indicating the effectiveness of dual task consistency for semi-supervised learning. Although these methods have the ability to mine effective information from unlabelled data to support task learning, they have no proper mechanism to exploit the cross-domain information, thus leading to limited segmentation results. Compared to these methods, Dual-Teacher leverages two teacher models to guide a student model for the learning of both intra-domain and inter-domain knowledge, thus achieving big improvements in terms of segmentation accuracy. Notably, our proposed AHDC obtains the best segmentation accuracy over these widely used and state-of-the-art semi-supervised methods, which shows its superiority for generalising across domains. Furthermore, it is observed that our proposed AHDC generally improves the segmentation accuracy compared to the joint training, which combines  the  cross-domain  data  directly for the semi-supervised LA segmentation. This demonstrates that our proposed AHDC can leverage cross-domain information to improve the model performance. We also provide qualitative comparison between different methods in Fig. \ref{fig:la_vis_comparison}. It is observed that the LAs estimated by other methods present fragmentary parts and unsmooth boundaries. While the LAs estimated by our proposed method are closer to the ground truth with smoother boundaries.

\subsection{Ablation Studies}
We performed ablation studies on $C1$ and $C2$ (C1 supports C2) to validate the effectiveness of our proposed AHDC for the cross-domain semi-supervised segmentation.

\begin{table}[!hbtp]
\captionsetup{justification=centering}
 \caption{\centering The performance of AHDC on different percentages of labelled data. Abbreviations: Lx (\%): Lx (\%): the ratio of labelled data in the training set of centre x;  Ux (\%): the ratio of unlabelled data in the training set of centre x; DSC, Dice Similarity  Coefficient;  JI, Jaccard Index; ASD, Average Surface Distance.}\label{table:ratio_label} 
 \centering
 \scalebox{.7}{
 \setlength{\floatsep}{10pt plus 3pt minus 2pt} 
\begin{tabular}{c|cc|ccc}
\addlinespace
\toprule
\multirow{2}*{Method} & \multicolumn{2}{c|}{Rate} & \multicolumn{3}{c}{Metrics}\\ \cline{2-6}
                      & L2/U2 (\%) & L1/U1 (\%) & DSC & JI  &ASD\\ \hline
Upper Bound & $100/0$    & $0/0$      & $0.926\pm0.021$    & $0.863\pm0.036$    & $0.867\pm0.46$ \\ \hline    
Baseline & $20/0$   & $0/0$      & $0.860\pm0.103$    & $0.765\pm0.122$    & $4.00\pm5.08$ \cr
AHDC     & $20/80$   & $0/100$   & $0.917\pm0.026$    &$0.848\pm0.043$     & $1.17\pm1.60$   \\ \hline
Baseline &$10/0$   & $0/0$       & $0.815\pm0.142$     & $0.706\pm0.153$    & $4.84\pm6.09$  \cr 
AHDC    & $10/90$   & $0/100$      &$0.891\pm0.039$     &$0.805\pm0.060$     &$1.98\pm2.65$    \\ \hline
Baseline & $5/0$   & $0/0$       &$0.776\pm0.134$     &$0.650\pm0.146$     &$6.51\pm5.52$   \cr 
AHDC    & $5/95$   & $0/100$       & $0.871\pm0.041$    &$0.773\pm0.060$     & $1.95\pm1.44$   \\ \midrule 
\end{tabular}
 }
\end{table}

\textbf{(1) Model variation study for bidirectional adversarial inference}: As the experimental results are summarised in TABLE \ref{table:bai_hcr},  the bidirectional adversarial inference with bidirectional reconstruction improves the LA segmentation accuracy in terms of DSC, JI and ASD compared with the BAI$_{wbr}/$ALI$/$BiGAN. The reason behind the improvements is that bidirectional reconstruction makes the relationship between matched samples specified and constrained. It guarantees that the matched samples are one-to-one correspondence for subsequent effective hierarchical dual consistency learning on cross-domain data. It is also observed that the segmentation accuracy is dropped while removing the skip connection from the domain mapping network. The reason behind this is that the domain mapping network (UNet structure) employs the skip connection to deliver the low-level information. It allows the samples adapted to another domain to maintain the same LA structures, which makes subsequent dual consistency learning effective. Furthermore, one can see that our proposed BAI has better performance for the downstream semi-supervised LA segmentation task compared to the fully adversarial ALICE method, which indicates the superiority of our proposed BAI.

\begin{table}[!hbtp]
\captionsetup{justification=centering}
 \caption{\centering Model variation study on C1 and C2 (C1 supports C2). Abbreviations: Lx (\%): the ratio of labelled data in the training set of centre x;  Ux (\%): the ratio of unlabelled data in the training set of centre x; DSC, Dice Similarity  Coefficient;  JI, Jaccard Index; ASD, Average Surface Distance.}\label{table:bai_hcr} 
 \centering
 \setlength{\floatsep}{10pt plus 3pt minus 2pt} 
\scalebox{.9}{
\begin{tabular}{c|ccc}
\addlinespace
\toprule
\multirow{2}*{Method} & \multicolumn{3}{|c}{Metrics}\\ \cline{2-4}
                      & DSC & JI  &ASD\\ \hline
Lower Bound                 & $0.860\pm0.103$    & $0.765\pm0.122$    & $4.00\pm5.08$ \\ \hline
BAI$_{eds}$ + HDC           &$0.879\pm0.039$   & $0.786\pm0.060$  &$1.59\pm0.875$ \cr      
BAI$_{wbr}$ + HDC           &$0.885\pm0.051$   & $0.798\pm0.076$  &$1.37\pm0.802$ \cr
ALICE + HDC                 &$0.896\pm0.033$   & $0.814\pm0.053$  &$1.42\pm1.09$ \cr
BAI + HDC$_{intra}$         &$0.893\pm0.048$ &$0.809 \pm 0.073$  &$1.90 \pm 2.81$  \cr
BAI + HDC$_{inter}$         &$0.900\pm0.045$  &$0.822\pm0.066$  &$1.51\pm1.24$   \cr
AHDC                        & $0.917\pm0.026$    &$0.848\pm0.043$     & $1.17\pm1.60$ \\ \midrule 
\end{tabular}
}
\end{table}

\textbf{(2) Model variation study for hierarchical dual consistency}: As the experiment results are summarised in TABLE \ref{table:bai_hcr}, the independent intra-domain and inter-domain dual consistency learning can both improve the LA segmentation accuracy compared to the lower-bound model. This indicates that the intra-domain dual consistency learning and the inter-domain dual consistency learning are effective to exploit the unlabelled data from cross-domain data. Furthermore, it is also observed that the intra-domain dual consistency and the inter-domain dual consistency can promote each other for the cross-domain semi-supervised segmentation. These results demonstrate the effectiveness of the hierarchical dual consistency for semi-supervised segmentation on cross-domain data.

\begin{figure}[!hbtp]
\begin{center}
\scalebox{.85}{
   \includegraphics[width=1\linewidth]{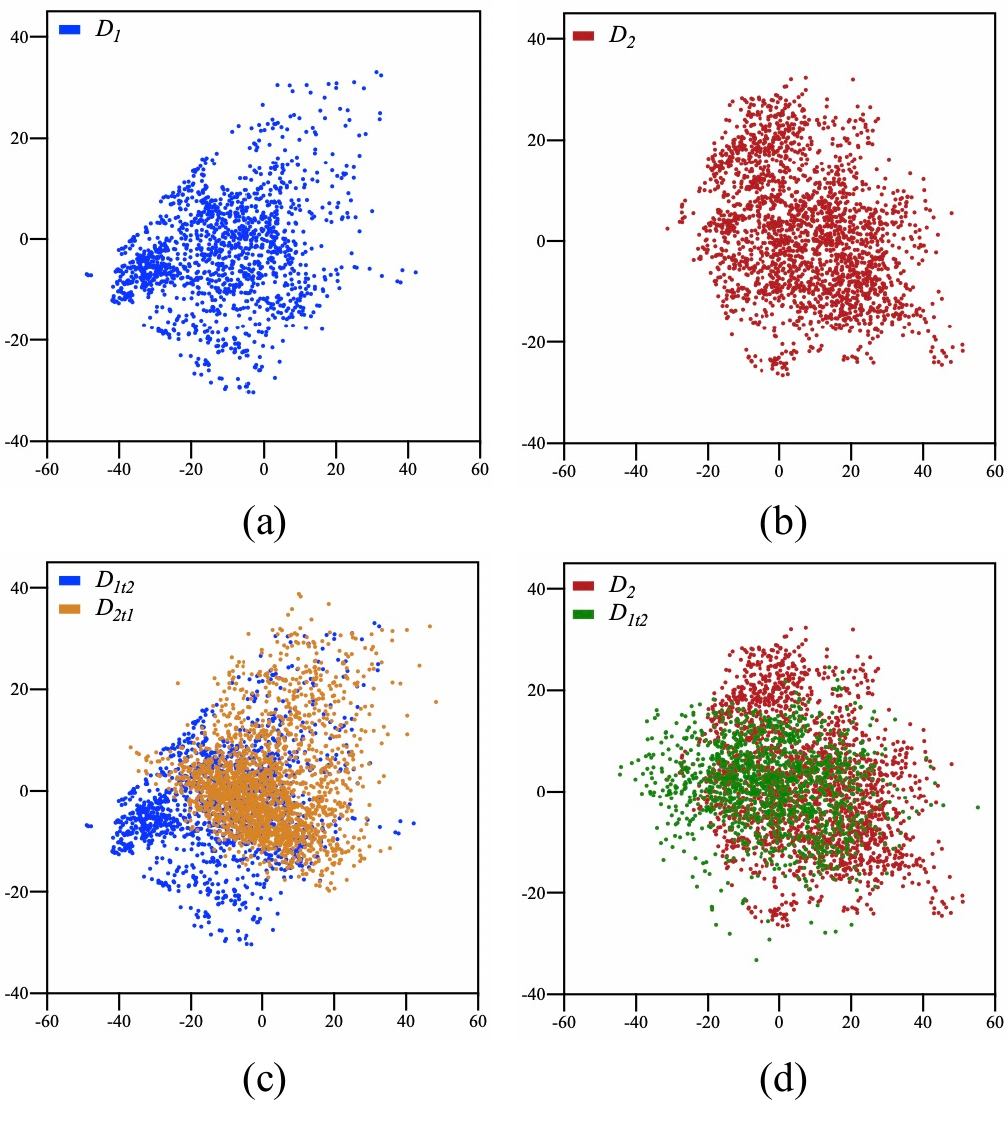}
   }
\end{center}
\caption{Principal components analysis based visualisation for the data distribution of the testing tests of C1 and C2. (a) The data distribution of domain $D_{1}$. (b) The data distribution of domain $D_{2}$. (c) The data distribution of domain $D_{1}$ and the domain $D_{2t1}$ adapted from $D_{2}$ to $D_{1}$. (d) The data distribution of domain $D_{2}$ and the domain $D_{1t2}$ adapted from $D_{1}$ to $D_{2}$.}
\label{fig:adaptation} 
\end{figure}

\subsection{The Effectiveness Analysis of BAI for Matching Domains}
The effectiveness of the bidirectional adversarial inference is further validated by the qualitative results on distribution alignment and sample matching in the testing set.

\textbf{(1) Distribution alignment}: In Fig. \ref{fig:adaptation}, we color samples from different domains and adapted domains to highlight their correspondence (brown and blue for the samples from domains of $D_{2}$ and $D_{1}$, respectively. Peru and green for the samples from the domain $D_{2t1}$ adapted from $D_{2}$ to $D_{1}$ and the domain $D_{1t2}$ adapted from $D_{1}$ to $D_{2}$, respectively). It is observed that the domains of $D_{1}$ and $D_{2}$ have different distributions as shown in Fig. \ref{fig:adaptation} (a) and Fig. \ref{fig:adaptation} (b). Besides, as shown in Fig. \ref{fig:adaptation} (c) and Fig. \ref{fig:adaptation} (d), after the bidirectional adversarial inference, the distribution of the domain $D_{2t1}$ adapted from $D_{2}$ to $D_{1}$ is consistent with the distribution of $D_{1}$. Meanwhile, the distribution of the domain $D_{1t2}$ adapted from $D_{1}$ to $D_{2}$ is consistent with the distribution of $D_{2}$. One also can find the adapted domains of $D_{1t2}$ and $D_{2t1}$ make the distribution spaces of $D_{1}$ and $D_{2}$ more complete. These results indicate the effectiveness of BAI for the distribution alignment. 
\begin{figure}[!hbtp]
\begin{center}
\scalebox{.9}{
   \includegraphics[width=1\linewidth]{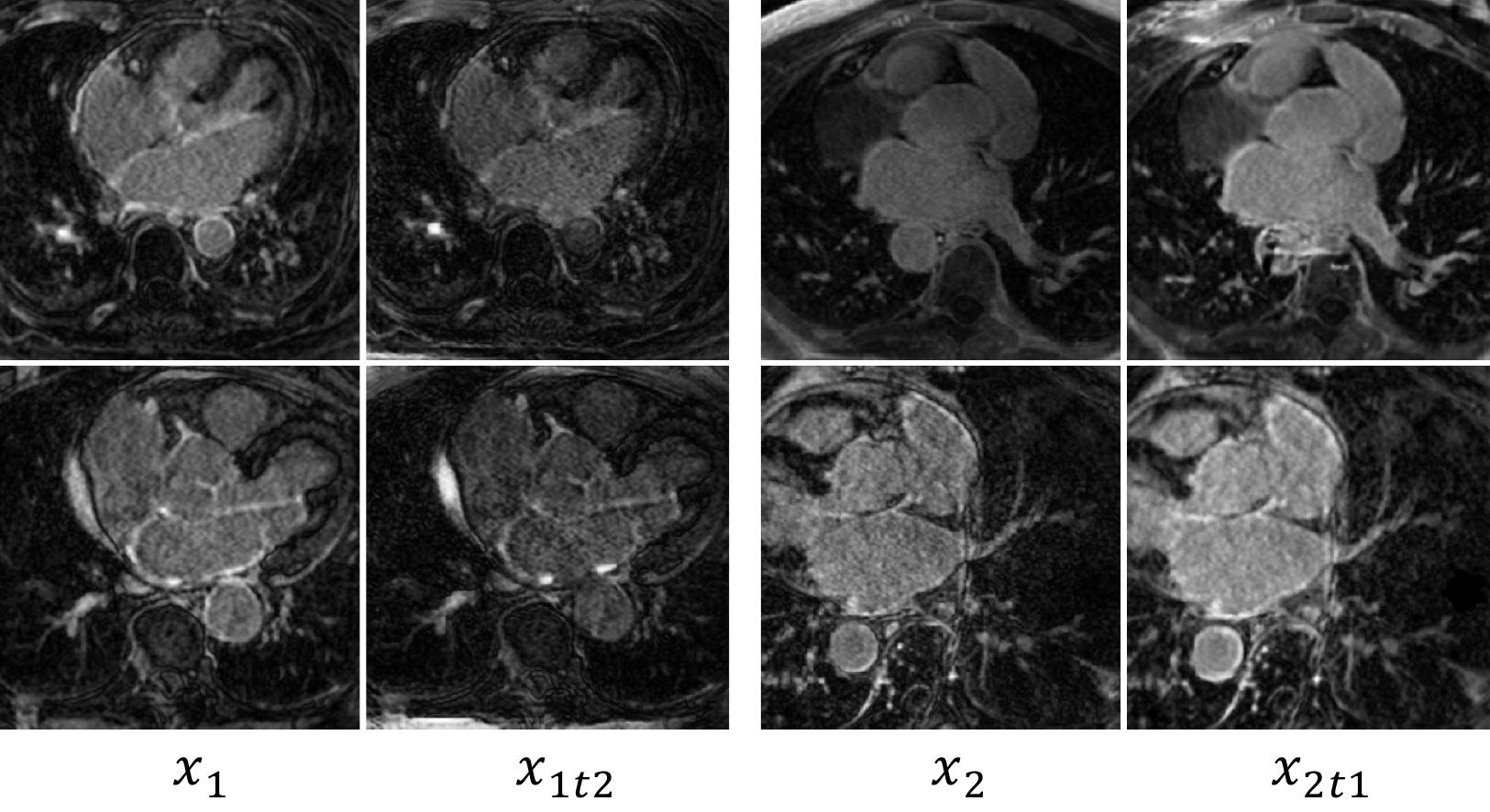}
   }
\end{center}
\caption{Qualitative visualisation of images and corresponding adapted images in the testing tests of C1 and C2. Abbreviations: $x_{1}$,\ image from domain $D_{1}$; $x_{1t2}$,\ image adapted from domain $D_{1}$ to domain $D_{2}$; $x_{2}$,\ image from domain $D_{2}$; $x_{2t1}$,\ image adapted from domain $D_{2}$ to domain $D_{1}$.}
\label{fig:adaptation_img} 
\end{figure}

\textbf{(2) Sample matching}: Fig. \ref{fig:adaptation_img} provides 2D visualisation of some examples before and after bidirectional adversarial inference. The images of the first two columns are from the domain $D_{1}$ and the domain adapted from $D_{1}$ to $D_{2}$. The images of the last two columns are from the domain $D_{2}$ and the domain adapted from $D_{2}$ to $D_{1}$. It is observed that the target shape and structure in corresponding images are consistent. However, the texture and the brightness in corresponding images are different. These results illustrate that the bidirectional adversarial inference is effective to produce the matched samples. 

\subsection{The Effectiveness Analysis of HDC for the Availability of  Complementary  Information}
\begin{figure}[!hbtp]
\begin{center}
\scalebox{.99}{
   \includegraphics[width=1\linewidth]{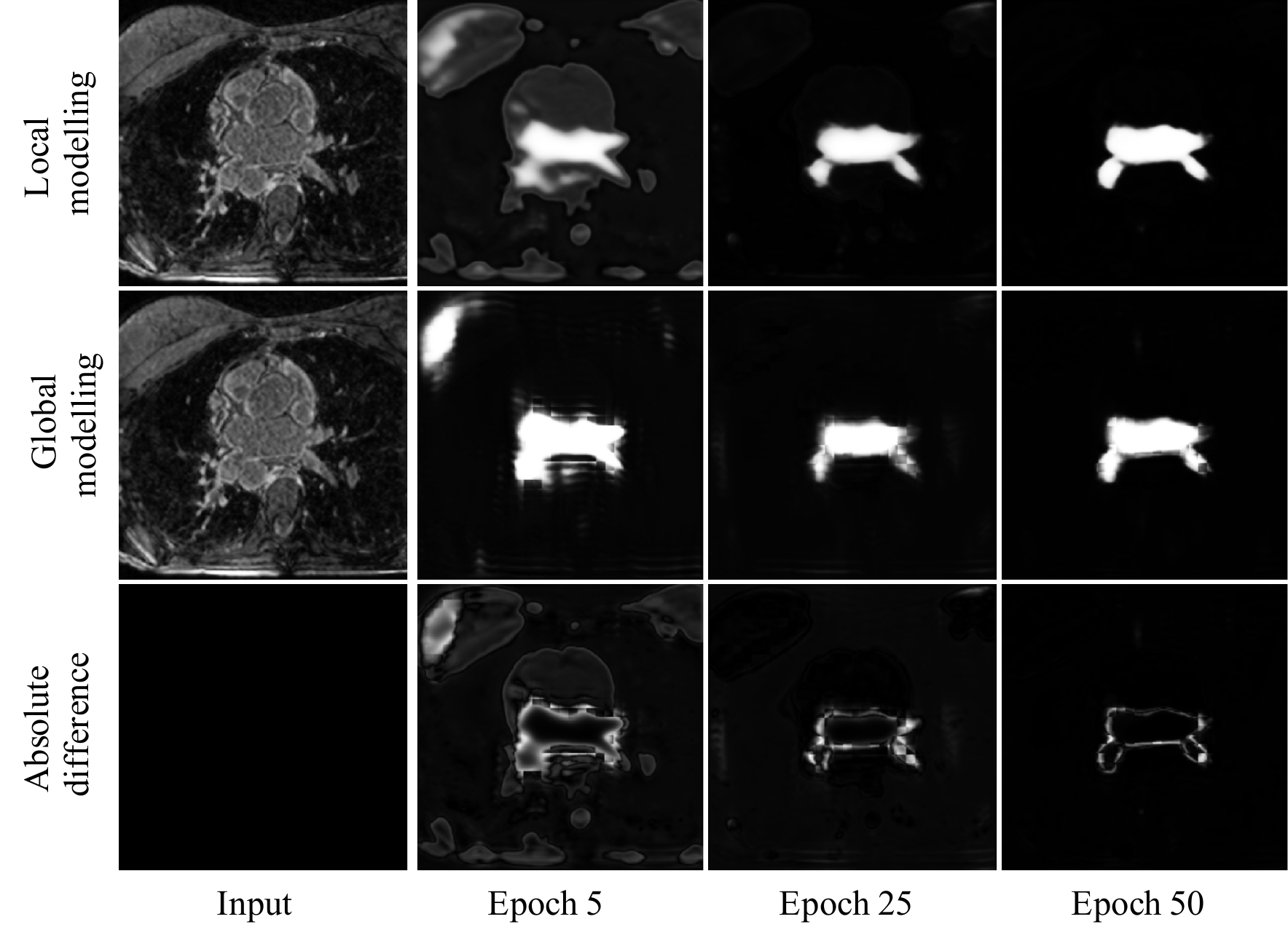}
   }
\end{center}
\caption{Visualization for the evolution of dual-modelling results (first row and second row) and their absolute difference (third row). The second to fourth columns correspond to the estimated LAs from 5, 25 and 50 epochs during model learning.}
\label{fig:local_global_difference} 
\end{figure}

\textbf{(1) Availability of the complementary modelling information}: TABLE \ref{table:local-global} summarises the experiment results on different modelling structures (Local-Global, Local-Local and Global-Global) for the intra-domain semi-supervised learning. It is observed that the dual-modelling structure (Local-Global) achieved higher segmentation accuracy. The reason behind this is that the dual modelling can complement each other during the model training, thus can provide effective prediction perturbation for consistency-based learning. We also visualize the examples estimated by local modelling branch and global modelling branch in different training epochs as shown in Fig. \ref{fig:local_global_difference}.  One can see that the absolute difference between the local modelling and global modelling demonstrates that the local modelling branch and local modelling branch are modelled separately, which can provide effective prediction perturbation for consistency based learning.

\begin{table}[!hbtp]
\captionsetup{justification=centering}
 \caption{\centering Performance comparison between dual structure (Local-Global) and non-dual structures (Local-Local and Global-Global) in terms of DCS, JI and ASD. The results are presented in the form of the mean (standard deviation). Abbreviations: DSC, Dice Similarity Coefficient; JI, Jaccard Index; ASD, Average Surface Distance; Local, local modelling network; Global, global modelling network.}\label{table:local-global} 
 \centering
 \setlength{\floatsep}{10pt plus 3pt minus 2pt}
\begin{tabular}{c|ccc}
\addlinespace
\toprule
\multirow{2}*{Method} & \multicolumn{3}{|c}{Metrics}\\ \cline{2-4}
                      & DSC & JI  &ASD\\ \hline
Local-Local     & $0.875\pm0.071$    & $0.785\pm0.102$    & $2.66\pm4.18$ \cr 
Global-Global   &$0.879\pm0.069$   & $0.789\pm0.098$  &$3.01\pm4.80$ \cr      
Local-Global    &$0.893\pm0.048$   & $0.809\pm0.073$  &$1.90\pm2.81$ \cr \hline
\end{tabular}
\end{table}

\begin{figure}[!hbtp]
\begin{center}
\scalebox{.9}{
   \includegraphics[width=1\linewidth]{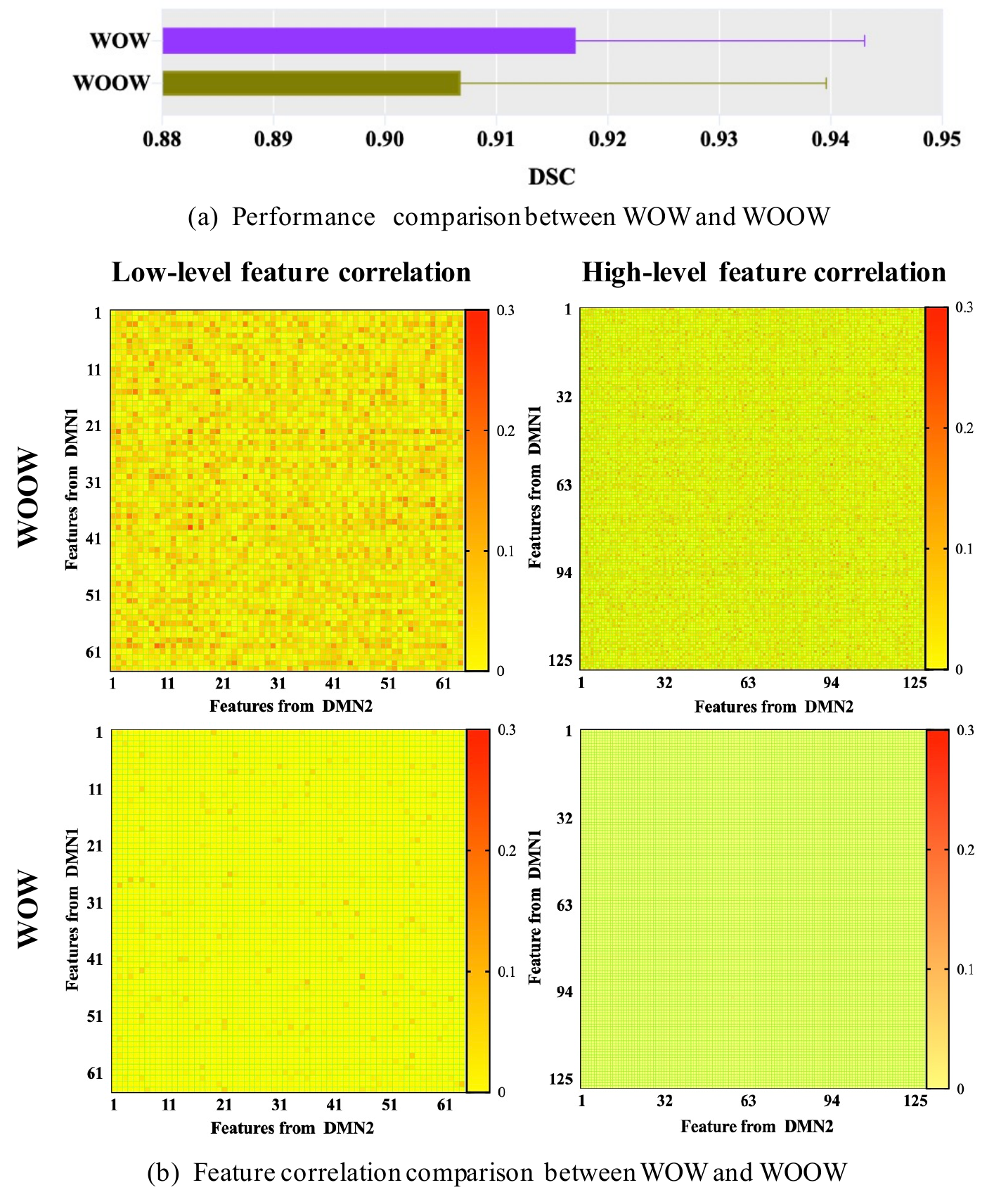}
   }
\end{center}
\caption{Segmentation performance comparison and feature correlation analysis between AHDC without orthogonal weights (WOOW) and AHDC with orthogonal weights (WOW). The experiments were performed on C1 and C2 (C1 supports C2). DMN1 and DMN2 represent two dual-modelling networks, respectively. The grape and asparagus bars denote the mean values with standard deviations.}
\label{fig:feature_correlation} 
\end{figure}

\textbf{(2) Availability of complementary domain information}: Fig. \ref{fig:feature_correlation} (a) provides the experiment results on hierarchical dual consistency learning with/without orthogonal weight constraint. Fig. \ref{fig:feature_correlation} (b) provides examples of feature correlations between corresponding layers of two dual-modelling networks with/without orthogonal weight constraint. It is observed that while removing the orthogonal weight constraint for inter-domain semi-supervised learning, the model segmentation performance is dropped. Meanwhile, the feature correlations between two dual-modelling networks become higher. The reason  behind this  is that  the inter-domain semi-supervised learning with the orthogonal weight constraint can provide more effective prediction perturbation for consistency based learning.  It is also observed that while removing the orthogonal weight constraint for inter-domain semi-supervised learning, the feature correlations between two dual-modelling networks are not high ($<0.3$). In this case, two dual-modelling networks also can learn the complementary domain knowledge for providing effective prediction perturbation, thus achieving a high segmentation accuracy of 0.907 in terms of DSC.

\subsection{The Effects of Parameter Settings on Model Performance}
TABLE \ref{table:parameter_validation} presents the performances of our model for the LA segmentation using different parameter settings. It is observed that our model achieves the best performance when the patch size and the $\lambda_{ow}$ are set as $8\times 8$ and $0.1$, respectively.

\begin{table}[!hbtp]
\captionsetup{justification=centering}
 \caption{\centering Parameter validation for AHDC framework. The results are presented in the form of mean $\pm$ standard deviation. Abbreviations: DSC, Dice Similarity Coefficient;  JI, Jaccard Index; ASD, Average Surface Distance.}
 \centering
 \scalebox{.9}{
    \begin{tabular}{ccccc}
    \toprule
    Parameter & \multicolumn{1}{c}{Value} & \multicolumn{1}{c}{DSC} & \multicolumn{1}{c}{JI} & \multicolumn{1}{c}{ASD (mm)}\\ \midrule
     \multirow{3}*{Patch Size} & $4\times 4$ & $0.874\pm0.062$ &$0.780 \pm 0.086$  &$2.12 \pm 2.22$\cr 
                               & $8\times 8$ &$0.893\pm0.048$   & $0.809\pm0.073$  &$1.90\pm2.81$\cr 
                               & $16\times 16$ & $0.881\pm0.069$ &$0.794 \pm 0.098$  &$2.40 \pm 3.85$\\ \midrule 
    \multirow{3}*{$\lambda_{ow}$} & 0.0 & $0.907\pm0.033$    &$0.831\pm0.052$     & $1.34\pm1.58$\cr 
                                  & 0.1 & $0.917\pm0.026$    &$0.848\pm0.043$     & $1.17\pm1.60$ \cr 
                                  & 1.0 & $0.913\pm0.032$    &$0.841\pm0.052$     & $1.34\pm1.68$\\ \midrule
    \end{tabular}
 }
 \label{table:parameter_validation} 
\end{table}

\section{Discussion}
In this study, we have developed a semi-supervised LA segmentation framework for generalising across domains. The semi-supervised LA segmentation framework comprises a BAI module and a HDC module. The effectiveness of each module has been validated in our ablation study presented in TABLE \ref{table:bai_hcr}. It is of note that self-attention based global modelling requires more computational resources, which are proportional to the dimensions of the image. In our proposed framework, we performed the self-attention based global modelling branch on the image feature maps for correlating $8\times 8$ patches instead of all pixels, which greatly reduces the requirements of computational resources during model training. Besides, during the testing phase or the practical applications, the self-attention based global modelling branches will be removed from our proposed framework. Then, the LA targets will be only predicted by the local-modelling branch with low computational resources.

\begin{table}[!hbtp]
\captionsetup{justification=centering}
 \caption{\centering Performance comparison between the vanilla model and our proposed AHDC using all the data available from four centres. Abbreviations: Lx ($\%$): the ratio of labelled data in the training set of centre x;  Ux ($\%$): the ratio of unlabelled data in the training set of centre x; DSC, Dice Similarity  Coefficient;  JI, Jaccard Index; ASD, Average Surface Distance.}\label{table:annotation_difference} 
 \centering
 \scalebox{.65}{
 \setlength{\floatsep}{10pt plus 3pt minus 2pt}
\begin{tabular}{c|cccc|ccc}
\addlinespace
\toprule
\multirow{2}*{Method} & \multicolumn{4}{c|}{Rate} & \multicolumn{3}{c}{Metrics}\\ \cline{2-8}
                      & \tabincell{c}{L1$/$U1 \\ ($\%$)} & \tabincell{c}{L2$/$U2 \\ ($\%$)} & \tabincell{c}{L3$/$U3 \\ ($\%$)} & \tabincell{c}{L4$/$U4 \\ ($\%$)} & DSC & JI  &ASD\\ \hline
\multirow{2}*{U-Net}  & $0/0$ & $100/0$ &$0/0$ &$0/0$  & $0.923\pm0.025$    & $0.858\pm0.042$    & $1.20\pm1.96$ \cr
                      & $100/0$  & $100/0$ & $100/0$ &$100/0$   &$0.927\pm0.022$   & $0.864\pm0.037$  &$0.728\pm0.578$ \cr \hline
                      
AHDC                  & $0/100$  & $100/0$ & $0/100$ & $0/100$ &$0.938\pm 0.015$   & $0.883\pm 0.026$  &$0.506\pm0.164$ \cr \midrule
\end{tabular}
 }
\end{table}

The AHDC requires the complementary domain information for inter-domain learning. For multi-centre studies, although the domains from different sources exhibit heterogeneous properties \cite{campello2021multi}, they still share some specific information because they come from the same image modality of LGE. To make the model focus on the heterogeneous properties of different domains for the inter-domain learning, we use an orthogonal weight constraint to extract the conditional independent features of different domains for subsequent target modelling. We have explored the effectiveness of the orthogonal weight constraint together with its weight coefficient $\lambda_{ow}$ for the inter-domain learning. As the experiment results are shown in TABLE \ref{table:parameter_validation}, one can see that the orthogonal weight constraint generally improves the segmentation accuracy. Furthermore, the performance of AHDC is not very sensitive to the $\lambda_{ow}$ values of 0.1 and 1.0 while using the orthogonal weight constraint. Therefore, the orthogonal weight constraint could exploit the heterogeneous properties among different domains for inter-domain learning.

Considering the data annotation scarcity in medical image analysis, our proposed method only requires the labelled data from one of the multiple centres during cross-domain learning, thus further reducing the dependence of the model on annotated data. As the experiment results are shown in TABLE \ref{fig:test} and TABLE \ref{table:MR_CT}, our proposed method is able to generalise across two different domains simultaneously. We further explore how the task model generalises across multiple domains. Specifically, we have applied our proposed method to the LGE CMRI data available from four centres.  We also trained a vanilla model (U-Net) with the LGE CMRI data available from a single target centre and all the LGE CMRI data available from four centres for comparison. As the experiment results are shown in TABLE \ref{table:annotation_difference}, compared with the results obtained by using all annotated data from a single domain, using all the data available from four centres only makes small improvements in the segmentation accuracy due to the domain shift and the label variations from different centres. While our proposed AHDC generally improves the segmentation accuracy, which indicates its ability for cross-domain semi-supervised learning.

\section{Conclusion}
In this paper, we proposed an adaptive hierarchical dual consistency for the cross-domain semi-supervised LA segmentation. The adaptive hierarchical dual consistency firstly overcomes the distribution difference and sample mismatch of different domains by the bidirectional adversarial inference. Then, it explores the complementary modelling and domain information in intra-domain and inter-domain for semi-supervised LA segmentation based on the hierarchical dual consistency. Comprehensive experiments on four 3D LGE CMR datasets and one CT dataset demonstrated the feasibility and superiority of our proposed method for the cross-domain semi-supervised LA segmentation.

\end{document}